\documentclass{aa}
\usepackage{natbib}
\bibpunct{(}{)}{;}{a}{}{,} % to follow the A&A style

 \usepackage[varg]{txfonts}
\usepackage{graphicx}
 \usepackage{lscape}
 
\usepackage{txfonts}
\usepackage{float} 
\usepackage{fourier}
\usepackage[overload]{empheq}
\usepackage[inline, shortlabels]{enumitem}

\begin{document} 

   \title{Center-to-limb variation of intensity and polarization in continuum spectra of FGK stars for
spherical atmospheres. }
\titlerunning{CLV of intensity and polarization in continuum spectra of FGK stars for 
spherical atmospheres.}
   \author{N. M. Kostogryz\inst{1}
          \and
          I. Milic \inst{2,3}
          \and
           S.V. Berdyugina \inst{1}
            \and
           P. H. Hauschildt \inst{4}
           }

   \institute{Kiepenheuer-Institut f\"ur Sonnenphysik (KIS), 
              Sch\"oneckstrasse 6, D-79104 Freiburg\\
              \email{kostogryz@kis.uni-freiburg.de, sveta@kis.uni-freiburg.de}
        	 \and
	   Max Planck Institute for solar system research (MPS), 
	   Justus-von-Liebig-Weg 3, D-37077 G\"ottingen
   \and
   Astronomical Observatory Belgrade, Volgina 7, 11060 Belgrade
    \and
    Hamburger Sternwarte, Gojenbergsweg 112, 21029 Hamburg
	     }

   \date{Received; accepted }

% \abstract{}{}{}{}{} 
 
  \abstract
   {}
   {One of the necessary parameters needed for the interpretation of the light curves of transiting exoplanets 
   or eclipsing binary stars, as well as interferometric measurements of a star or microlensing events 
   is how the intensity and polarization of a light change from the center to the limb of a star. 
   Scattering and absorption processes in stellar atmosphere affect both the center-to limb 
   variation of intensity (CLVI) and polarization (CLVP). 
   In this paper, we present a study of the CLVI and CLVP in continuum spectra considering different contributions of 
   scattering and absorption opacity for different spectral type stars with spherical atmospheres. }
   {We solve the radiative transfer equation for polarized light in the presence of continuum scattering,
   considering spherical model of a stellar atmosphere. We developed two independent codes based on 	
    Feautrier and short characteristics methods, respectively, to cross-check our results.
   }
   % results heading
   {We calculate the center-to-limb variation of intensity (CLVI) and 
polarization (CLVP) in continuum for the Phoenix grid of spherical stellar model atmospheres 
for a range of effective temperatures ($4000 - 7000 \rm K$), gravities ($\log g = 1.0 - 5.5$) 
and wavelengths ($4000 - 7000 \rm \AA$), which are tabulated and available at the CDS. 
In addition, we present several tests of our codes and compare
our calculations for the solar atmosphere with published photometric and 
polarimetric measurements. We also show that our two codes provide similar results in all considered cases.
   }
   {For sub-giant and dwarf stars ($\log g = 3.0 - 4.5$), lower gravity and lower effective 
   temperature of a star lead to higher limb polarization of the star.
   For giant and supergiant stars ($\log g = 1.0 - 2.5$), the highest effective temperature yields  the largest
   polarization. By decreasing of the effective temperature of a star down to $4500 - 5500 \rm K$ 
   (depending on $\log g$) the limb polarization decreases and reaches a local minimum. 
   It increases again down to temperatures of $4000 \rm K$. 
   For the most compact dwarf stars ($\log g = 5.0 - 5.5$) the limb polarization degree shows a maximum 
   for models with effective temperatures in the range $4200 - 4600 \rm K$ (depending on $\log g$)
   and decreases toward higher and lower temperatures.
   }
   

   \maketitle
%
%________________________________________________________________

\section{Introduction}
Center-to-limb variation (i.e., the angular variation) of intensity (CLVI) and polarization (CLVP) are necessary for interpretation of 
light curves of a star during exoplanetary transits \citep[e.g.,][]{muller13, kostogryz15b} of eclipsing binary systems \citep[e.g.,][]{bass12, kemp83} 
as well as for interpreting interferometric observations
\citep[e.g.,][]{wittkowski04, chiavassa10} and microlensing observations \citep[e.g.,][]{an02},
and helioseismology \citep{kuhn97, toutain99, kuhn12}. 

Proper interpretation of exoplanet light curves resulting from high-precision photometry (e.g. 
obtained by Kepler mission) requires precise computations of center-to-limb variations of the
intensity (i.e. limb darkening/brightening) for stars of different spectral classes.
Limb darkening calculations for extensive grids of plane-parallel 1D hydrostatic stellar model atmospheres were made by different authors (see, e.g., \cite{claret00, sing10}).
In our previous works, we studied the CLV of intensity and polarization
for plane-parallel stellar atmosphere models of different spectral classes (FGK) and surface gravities $\log g = 3.0 - 5.0$ \citep{kostogryz15a}. 
Recently, \cite{harrington15} calculated linear polarization in continuum for MARC stellar atmosphere models \citep{gustafsson08} and
found it systematically higher than ours for the same effective temperatures
and log g. It was suggested that this difference results from contribution of spectral lines. Spectral lines are important 
for blue wavelengths where there may be little real continuum observed, but for $4000 \AA$ and longer the difference in our results can be due to
considering different stellar model atmosphere
(see, for example, comparison of continuum polarization for several solar models by \cite{kostogryz15a}).
We will investigate line contribution in a separate paper.

However, for lines of sight near the limb of the star, the approximation of plane-parallel model 
atmosphere is not valid. The reason for that is that these 'rays' take significantly different 
path from what is predicted by a plane-parallel assumption. The more extended the stellar atmosphere
(lower surface gravity) is, the more pronounced this effect is. Therefore, to obtain precise values of 
intensity near the limb of the star one needs to calculate CLVI using spherical model atmospheres. 
There are several theoretical studies where limb darkening coefficients in different photometric 
bands are computed for stars with spherical model atmospheres \citep[see, for example][]{claret12, 
claret13, neilson13a, neilson13b}.
 
Consider, for example, an exoplanetary transit. As the ingress and egress moments are critical 
for describing the light curves, accurate calculations, especially close to the limb, are necessary 
for the interpretation of light curves. In some cases it is not sufficient  to use polynomial fit in $\mu$
to describe stellar limb darkening.

On the other hand,  the center-to-limb variation of linear polarization in the continuum is not so 
well studied despite the fact that it yields a lot of information about stellar atmospheres.
Recently, \cite{kostogryz15b} showed that variation of linear polarization during a transit provides
information on physical parameters of stellar spots (such as size and position), as well as 
on the orbital parameters of exoplanets (for example, longitude of ascending node that
cannot be derived from the photometric light curve). The crucial input in these calculations is the center-to-limb 
variation of linear polarization. 
   
In this paper we calculate the CLVI and CLVP in continuum spectra 
of different spectral classes under the assumption of spherical geometry. In addition, we expanded
the range of effective temperatures and surface gravities as compared to our previous study \citep{kostogryz15b}.
As there are no earlier results for 
CLVP for spherical atmosphere of different stars to compare our calculations with, we developed two 
independent codes for solving the radiative transfer equation for polarized light in presence of continuum scattering, 
in order to make a cross-check of our 
results. In the next section we describe the two methods of solving the radiative 
transfer equations in detail. Section 3 contains the comparative tests and results of our calculations of
solar and stellar CLVI and CLVP. The conclusions of our paper are presented in the fourth section.

\section{Methods of calculation}

Polarized radiation is fully described by four Stokes parameters $I, Q, U$ and $V$. 
It is customary to assume a coordinate system where Stokes parameter $Q$ is either 
parallel or perpendicular to the local limb \citep[here we choose the former, as in][]{kostogryz15a}. 
In this paper we are considering continuum polarization due 
to scattering by atoms and free electrons (Rayleigh and Thomson scattering, respectively). 
This means that,  as in one-dimensional models there is no way of breaking 
axial symmetry, Stokes $U$ is equal to zero. The same is valid for the $V$ component, 
as neither scattering nor magnetic fields can produce circular polarization in the continuum. Note that in 
case of spectral lines both Stokes $U$ and $V$ can be non-zero, via mechanisms 
known as Hanle and Zeeman effect, respectively \citep[see, for example, monograph by][]{LL04}.

Therefore, in the problem we are facing, the Stokes vector can be described as follows:
\begin{equation}
	\bold{I} (r, \mu, \lambda)~ = ~ \left( \begin{array}{c}
	I \\
	Q  \end{array} \right), 
\end{equation}
where $\mu = \cos\theta$ and $\theta$ defines a line of sight direction with respect to the atmospheric 
normal. For the direction corresponding to the center of stellar disk $\mu = 1$, while for the stellar limb $\mu = 0$. 
$r$ is the radial coordinate (our models are spherically 
symmetric and hence one-dimensional) and $\lambda$ is the wavelength of the radiation. 

\begin{figure}
	\includegraphics[width = 0.49\textwidth]{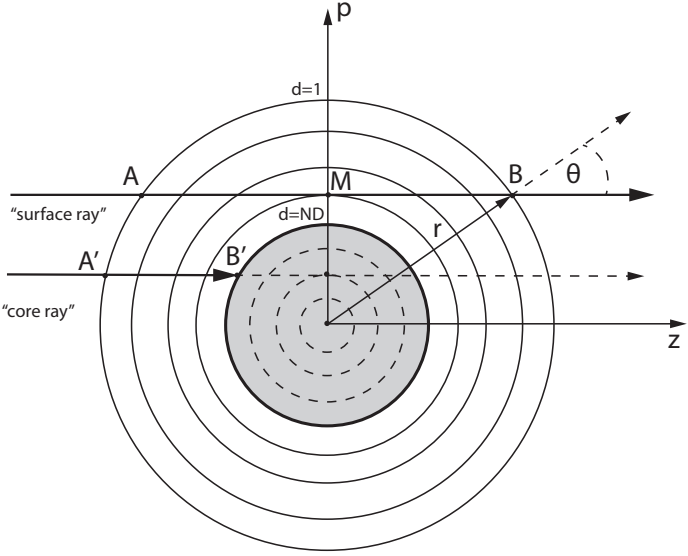}
	\caption{Description of $p-z$ coordinate system that is employed for the
	solution of radiative transfer equation of polarized light. 
}
\label{sketch}
\end{figure}

To obtain the emergent Stokes vector, one has to solve the polarized radiative transfer equation in 
the stationary spherically symmetric media \citep[e.g.\,][]{SA3}:
 \begin{equation}
	 \mu~\frac{\partial\bold{I} (r, \mu, \lambda)}{\partial r} +  \frac{1-\mu^2}{r} 
	\frac{\partial \bold{I} (r, \mu, \lambda)}{\partial \mu}= ~-~(k_c+\sigma_c)~\bold{I} 
	(r, \mu, \lambda)~+~\bold{\eta} (r, \mu, \lambda), 
 \label{rte_spherical}
\end{equation}
where $\bold{\eta}$ is emission term which comes from both the thermal emission 
(free-free and free-bound processes) and from scattering processes. $k_c$ and $\sigma_c$
stand for coefficients of absorption and scattering, respectively. Their sum gives total absorption koefficient.
It is common to refer to the ratio of emission and absorption coefficients as 
the "source function". The source function ($\bold{S}$) in the continuum can be written as:
 \begin{equation}
 	{\bold{S} (r, \mu, \lambda)}~ = 
	~\frac{1}{(k_c~+~\sigma_c)}~(k_c~\bold{B}(\lambda)~+~\sigma_c~\bold{S}_{s} (r, \mu, 
	\lambda)), 
 \label{full_source}
 \end{equation}
 where $\bold{B}(\lambda)$ describes photon creation from the thermal pool of the gas 
and is described by the Planck function:

 \begin{equation}
 	\bold{B}(\lambda)~ = ~ \left( \begin{array}{c}
	B (\lambda, T) \\
	0  \end{array} \right), 
 \end{equation}
$\bold{S}_{s}(r, \mu, \lambda)$  defines the contribution from scattering 
sources, and is given by:
 \begin{equation}
 	\bold{S}_{s}~(r, \mu, \lambda)~ = ~\int_{-1}^{1} \bold{P}_R(\mu, \mu 
	')~\bold{I}_\lambda(r, \mu ', \lambda)~\frac{d\mu '}{2}, 
 \label{source}
 \end{equation}
 where $\mu '$ is the direction of the incident radiation, and 
 $\bold{P}_R$  is the Rayleigh phase matrix that takes the angular dependence of 
 Rayleigh and Thomson scattering \citep[e.g.,][]{stenflo94}:
 \begin{equation}
 	\bold{P}_{R}~ = ~ \bold{E}_{11}~+~\frac{3}{4}~\bold{P}^2.
 \end{equation}
Matrix $\bold{E}_{11}$ has only one non-zero element, $E_{11}=1$ that describes unpolarized, isotropic scattering.
The matrix $\bold{P}^2$ describes the linear polarization scattering and has the following form:
 \begin{equation}
	\bold{P}^2~ = ~\frac{1}{2}~ \left( \begin{array}{cc}
	\frac{1}{3}(1-3\mu^2)(1-3\mu '^2)  &  (1-3\mu^2)(1-\mu '^2)\\
	(1-\mu^2)(1-3\mu '^2) & 3(1-\mu^2)(1-\mu '^2)  \end{array} \right).
 \end{equation}
To solve the radiative transfer equation in a spherical coordinate system (Eq. \ref{rte_spherical}), 
it is advantageous to describe the angular and spatial dependence of the 
radiation field by a set of parallel rays that are tangential to the discrete 
radial shells. This is known as the "along-the-ray" approach \citep{avrettloeser84}. 
The new coordinate system is described by the impact parameter $p$ that illustrates rays and 
the distance along $z$ (Fig.\ref{sketch}).
The radius in the new coordinate system can be written through the $p$ and $z$ as follows: 
\begin{equation}
	r~=~\sqrt{p^2+z^2}. 
\end{equation}
Each ray $p$ intersects the radial shells $r$ at an angle $\theta$ whose cosine $\mu$ is written as
\begin{equation}
	\mu~=~\frac{\sqrt{r^2-p^2}}{r}~=~\frac{z}{\sqrt{p^2+ z^2}}. 
\end{equation}
Then the differential operator in $z$ is given by
\begin{equation}
	\frac{\partial}{\partial z} = \mu \frac{\partial}{\partial r} + 
	\frac{1-\mu^2}{r} \frac{\partial}{\partial \mu}.
\end{equation}
It is common to introduce a new variable, the optical depth, defined as
\begin{equation}
	d\tau(z, p, \lambda) = - (k_c+\sigma_c)~dz.
\end{equation}
After all transformations we have spherical radiative transfer equation of the rays 
travelling along $\pm z$ for each ray with its impact parameter $p$ as follows:
\begin{equation}
 	\pm \frac{\partial \bold{I}^\pm(z, p, \lambda)}{\partial \tau(z, p, \lambda)}~=~ 
	~\bold{I^\pm} (z, p, \lambda)~-~\bold{S} (z, p, \lambda).
 \label{rte}
\end{equation}
It is important to note that equations \ref{full_source}, \ref{source} and \ref{rte} 
are coupled. Substituting Eq.\,\ref{source} into Eq.\,\ref{full_source} and then in Eq.\,\ref{rte} 
would lead to an integro-differential equation for the specific intensity and polarization. With given boundary 
conditions, this integro-differential equation can be solved directly but it turns out that the 
iterative solution of the coupled systems of equations \ref{full_source}, \ref{source} and 
\ref{rte} is faster and more robust. Starting from the initial guess for the source function 
(usually $S^I=B$ and $S^Q = 0$) one computes angular and spatial distribution of the intensity and 
polarization and uses them to compute new value of the source functions. The process is then repeated and 
iterated until convergence. This approach is known as $\Lambda$ iteration. 
For a much wider and more detailed insight in the problem, see, for example, new edition of the 
classical reference ``Stellar Atmospheres'', by \citet{SA3}.

In this work we use the iterative approach to the problem. The main issue then is to 
numerically solve Eq.\,\ref{rte}, given the current values of the source function. 
This equation is known as the formal solution. We perform the formal solution using 
two different techniques that are described in the next two subsections. 

\subsection {Feautrier solution of the radiative transfer equation}

The first approach to the numerical solution of the radiative transfer equation considered 
in our paper is the Feautrier method modified for a spherically symmetric coordinate system
\citep[][etc]{humryb71, mihalas78, senwilson98, hoffmann14}. In this method the 
two radiative transfer equations (one for I and one for Q)(Eq.\ref{rte}) can be rewritten  
as a two-point boundary problem in the form of second order differential equations. 
Defining the mean-intensity-like and flux-like variables as follows
\begin{align}[left = \empheqlbrace]
	\bold u(z, p, \lambda)~&=~\frac{\bold{I}^+(z, p, \lambda)+\bold{I}^-(z, p, \lambda)}{2} 
	\nonumber \\ 
	\bold v(z, p, \lambda)~&=~\frac{\bold{I}^+(z, p, \lambda)-\bold{I}^-(z, p, \lambda)}{2}. 
\label{uvdef}
\end{align}

Note that $\bold u$ and $\bold v$ are vector quantities given by $\left( 
\begin{array}{c}
	u^I \\
	u^Q  \end{array} \right) $,
and $ ~ \left( \begin{array}{c}
	v^I \\
	v^Q  \end{array} \right) $, 
respectively. 
Substituting eq. \ref{uvdef} in eq. \ref{rte}, the later can be rewritten as two differential equations: 
\begin{align}[left = \empheqlbrace]
	\frac{\partial\bold {u}(z, p, \lambda)}{\partial\tau(z, p, \lambda)}~&=~\bold{v}(z, p, 
	\lambda) \nonumber \\ 
 	\frac{\partial\bold {v}(z, p, \lambda)}{\partial\tau(z, p, \lambda)}~&=~\bold{u}(z, p, 
	\lambda) -\bold{S}(z, p, \lambda).
\end{align}

The combination of these two equations yields the second order differential radiative transfer equation:
\begin{equation}
	\frac{\partial^2\bold {u}(z, p, \lambda)}{\partial\tau(z, p, \lambda)^2}~=~\bold{u}(z, 
	p, \lambda) -\bold{S}(z, p, \lambda).
\label{rte_F}
\end{equation}

To solve the differential equation one needs boundary conditions.
The upper boundary condition at $r = r_1$ (see Fig. \ref{sketch}) where we assume no incident radiation 
$(\bold{I^-} \equiv 0)$ is
\begin{equation}
	\bold{u}~=~\bold{v},
\end{equation}
which, further, results in the following expression at the boundary condition $z=z_{max}$:
\begin{equation}
	\frac{\partial\bold {u}(z, p, \lambda)}{\partial\tau(z, p, 
	\lambda)}\Biggr\rvert_{\tau(z_{max}, p, \lambda)=0}~=~\bold{u}(z_{max}, p, \lambda),
\label{upbc}
\end{equation}
where $z_{max} = \sqrt{r_1^2 - p^2}$.

At the lower boundary two cases should be distinguished: the rays that intersect 
the core ($p \le r_{ND}$, "core rays") from
those that do not ($p>r_{ND}$, "surface rays") (see Fig. \ref{sketch}). For the core rays, we assume that 
radiation coming up from the bottom of the star is not polarized ($Q^+ = 0$), and we use the diffusion 
approximation for the intensity $I^+(\mu) = S + \mu \frac{\partial S}{\partial \tau}$. For the surface rays 
we use a "reflecting" boundary $\bold{I}^+ = \bold{I}^-$ both for intensity and polarization: 
\begin{align}[left = \empheqlbrace]
	\frac{\partial{u^I}(z, p, \lambda)}{\partial\tau(z, p, 
	\lambda)}\Biggr\rvert_{\tau_{max}}~&=~I_{core}-u^I(z, p, \lambda) \nonumber \\
	\frac{\partial{u^Q}(z, p, \lambda)}{\partial\tau(z, p, 
	\lambda)}\Biggr\rvert_{\tau_{max}}~&=~u^Q(z, p, \lambda)
\label{corebc}
\end{align}
for the core rays, and 
\begin{align}[left = \empheqlbrace]
	\frac{\partial{u^I}(z, p, \lambda)}{\partial\tau(z, p, 
	\lambda)}\Biggr\rvert_{\tau_{max}}~&=~0\nonumber \\
	\frac{\partial{u^Q}(z, p, \lambda)}{\partial\tau(z, p, 
	\lambda)}\Biggr\rvert_{\tau_{max}}~&=~0
\label{non_corebc}
\end{align}
for the surface rays. 

Therefore, we numerically solve the radiative transfer equation Eq.\ref{rte_F} with boundary conditions Eq. \ref{upbc}, \ref{corebc}, \ref{non_corebc},
which, after making the variables discrete, leads to a block tri-diagonal system. More details on the solution of radiative transfer equation in spherical coordinates can be found in \cite{peraiah01}.

\subsection{Short characteristics solution}

Short characteristics method for the numerical formal solution \citep{KA_SC87, MAM78} 
is based on the integral form of the radiative transfer equation, i.e. on the ``formal'' solution 
on the line segment connecting appropriately chosen \emph{upwind point} U and \emph{local point} L:
\begin{equation}
	\bold{I}_{\rm L} =\bold{I}_{\rm U} e^{-\Delta} + \int_0^{\Delta} \bold{S}(t) 
	e^{-t} dt,
 \label{SC1}
\end{equation}
where $\bold{I}_{\rm L}$ and $\bold{I}_{\rm U}$ are specific monochromatic intensities in the local 
and upwind point, respectively, $\Delta$ is monochromatic optical path between the 
points and $\bold{S}$ is the source function. Now, to numerically solve the integral in Eq.\,\ref{SC1}, 
one usually assumes either low-order polynomial or spline behaviour of the source 
function on the interval UL, which leads to the following scheme for the formal solution:
\begin{equation}
 	\bold{I}_{\rm L} = \bold{I}_{\rm U} e^{-\Delta} + w_{\rm U} \bold{S}_{\rm U} + 
	w_{\rm L} \bold{S}_{\rm L} + w_{\rm L'} \bold{S}'_{\rm L},
 \label{SC2}
\end{equation}
where $w$ are weights which depend on $\Delta$, and $\bold{S}'_{\rm L}$ is the derivative of 
the source function in the local point along the direction of the proparagation of radiation 
(i.e. `along-the-ray'). This derivative is evaluated either explicitly or expressed through 
$\bold{S}_{\rm L}$, $\bold{S}_{\rm U}$ and, in some schemes by means of an additional point, D, 
in the downwind direction of the ray propagation. In one-dimensional schemes, upwind 
and downwind points are chosen to be the previous and following grid points, with 
respect to the direction of propagation of the radiation. Originally, first or second order 
polynomials have been used to obtain weights in Eq.\,\ref{SC2}, but strictly monotonic 
Bezier splines  lead to a better behaved numerical solution \citep{Auer03, Jaime13}. 
In the computations presented in this paper we have used Bezier splines of the second 
order. In the case of the linearly polarized radiation we are considering here, the formal 
solution for the Stokes vector is obtained by performing the formal solution for $I$ and $Q$ 
components separately. To recap this brief description, it is important to remember that, 
to obtain the specific intensity in the local point, one needs appropriate intensity in the 
upwind point and values of emissivity and opacity in upwind, local, and, depending on 
the numerical scheme, possibly downwind point. 

As noted in the previous subsection, radiative transfer in spherical media is simplified 
by the use of the so called along-the-ray approach \citep{avrettloeser84}. In this 
approach, short characteristics method is straightforward to apply. To obtain the full radiation 
field, the radiative transfer equation is solved ray-by-ray where rays are described by the impact 
parameter $p$. Consider the ray described by the line segment AB in Fig.\,\ref{sketch}: 
First notice that to formally solve the radiative transfer equations on segment AB is, because of the symmetry arguments, 
identical to solving the equation on segment AM in the direction A$\rightarrow$M, with the
boundary condition $I_{\rm A} = 0$, followed by the solution in the opposite direction, 
with the boundary condition $I^{\leftarrow}_{\rm M} = I^{\rightarrow}_{\rm M}$. This is identical
to treating the segment AM as a single, one-dimensional, plane-parallel atmosphere with 
two rays propagating in directions $\mu = -1$ and $\mu = 1$. The procedure for the ray 
A'B' is similar, except the boundary condition in B' is different and follows from the diffusion 
approximation (Stokes $I$ component), or is equal to zero (Stokes $Q$ component). 

After the intensity has been computed at all points and for each ray, we can compute the source 
function using Eq.\,5. The process is repeated until convergence. A standart way of 
accelerating this convergence is to use Jacobi iteration, also known as `operator splitting', 
`operator perturbation' or Accelerated Lambda Iteration (ALI). In the cases considered here 
it is not necessary as the optical thickness of the medium is rather moderate but in general it 
leads to an increase in convergence by a order of magnitude. For a review of ALI see, for 
example, \citet{Hubeny03}. 

In principle, the only difference between the two approaches we have used is in the formal 
solution (Feautrier versus the short characteristics), which is, again, the most important part 
of the computation as the angular integration of the specific intensity in order to obtain the 
source function is rather straightforward. We now describe the stellar models we have used 
and discuss the main contributors to the continuum opacity.

\subsection{Stellar models and opacities}

Emergent polarized intensity is uniquely determined by the atmospheric model, i.e. the change of 
temperature and pressure with geometrical height in the atmosphere.  We use spherically
symmetric PHOENIX local thermodynamic equilibrium model atmospheres \citep{husser13}. 
We consider effective temperatures in the range $4000 \le T_{\rm eff} \le 7000 \rm K$ in steps of $100 \rm K$ 
and surface gravities $1.0 \le \log g \le 5.5$ in steps of 0.5. The metallicity is chosen to be equal to the solar for all models.

\begin{figure*}
	\centering
	\begin{minipage}{0.9\textwidth}
	  	\centering
		\includegraphics[width=.45\linewidth]{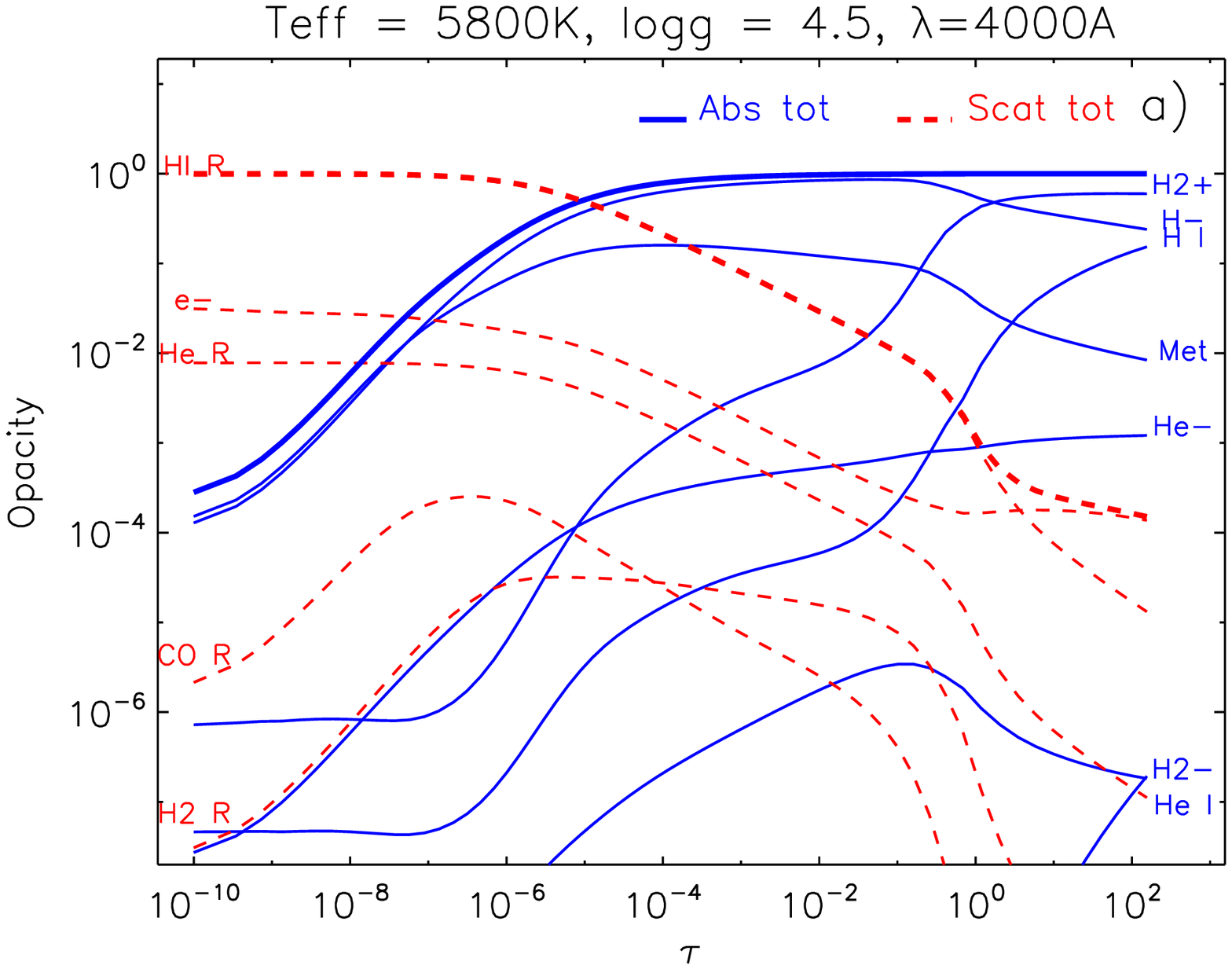}
		\includegraphics[width=.45\linewidth]{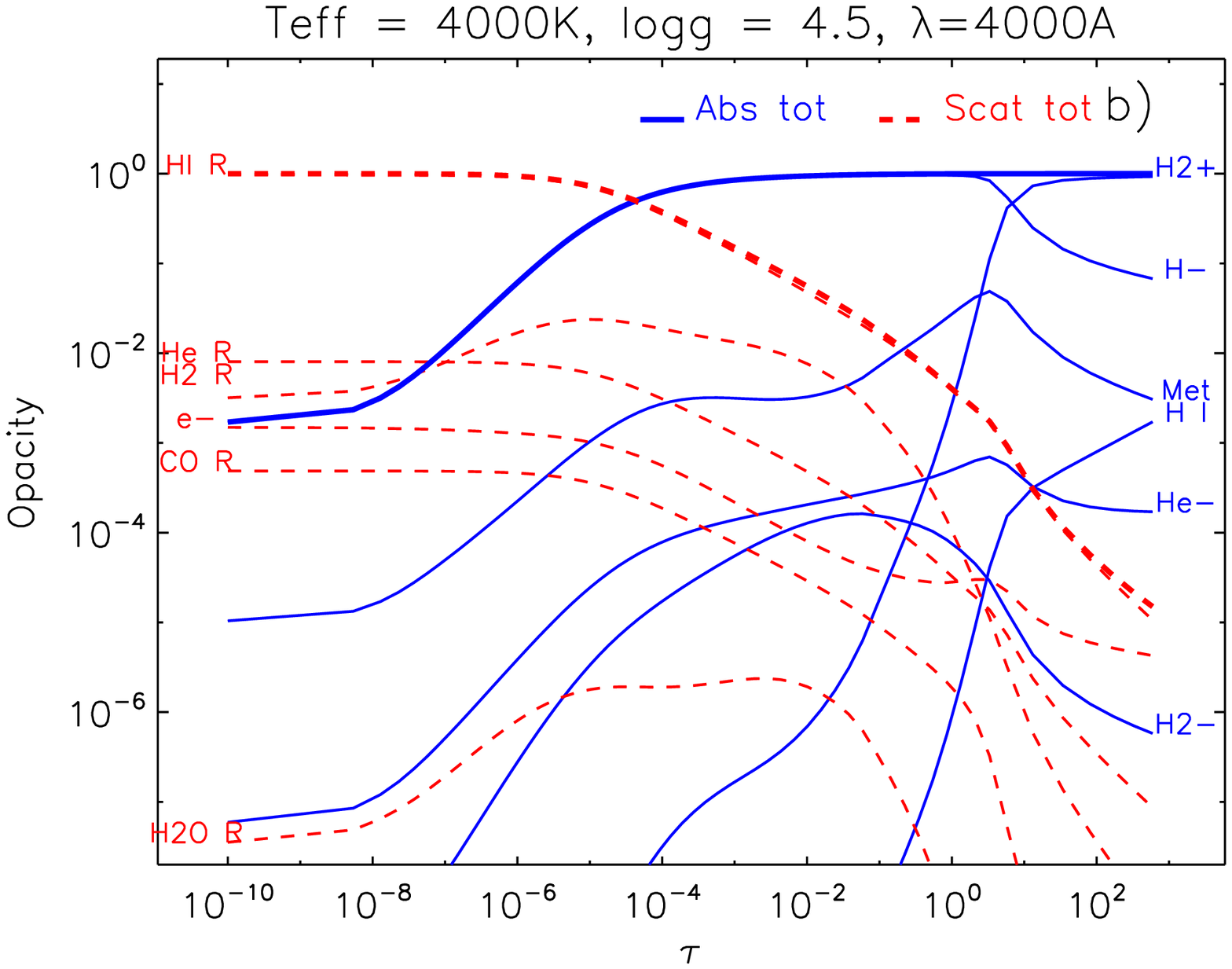}
	\end{minipage}

	\begin{minipage}{0.9\textwidth}
 		\centering
		\includegraphics[width=.45\linewidth]{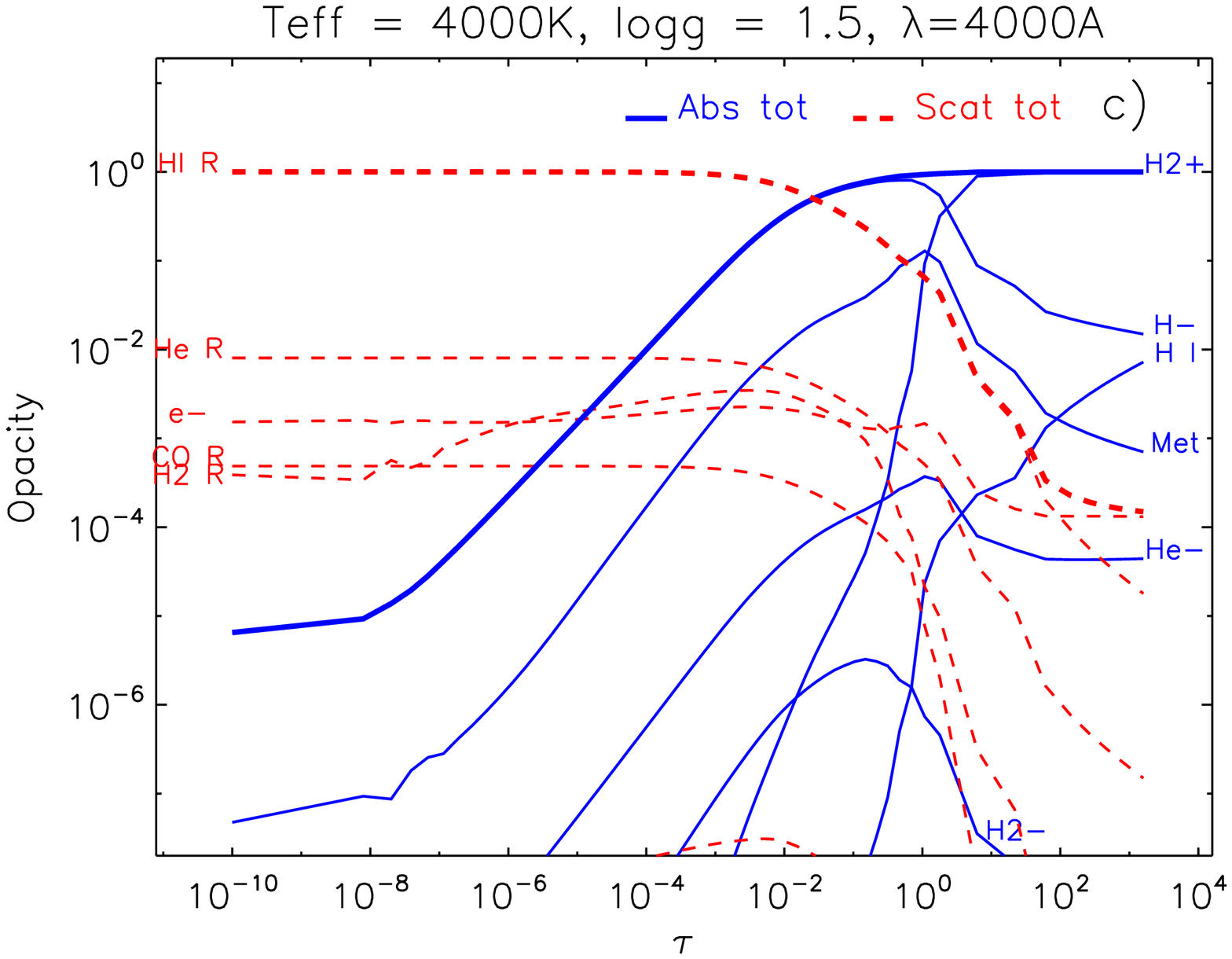}   
		\includegraphics[width=.45\linewidth]{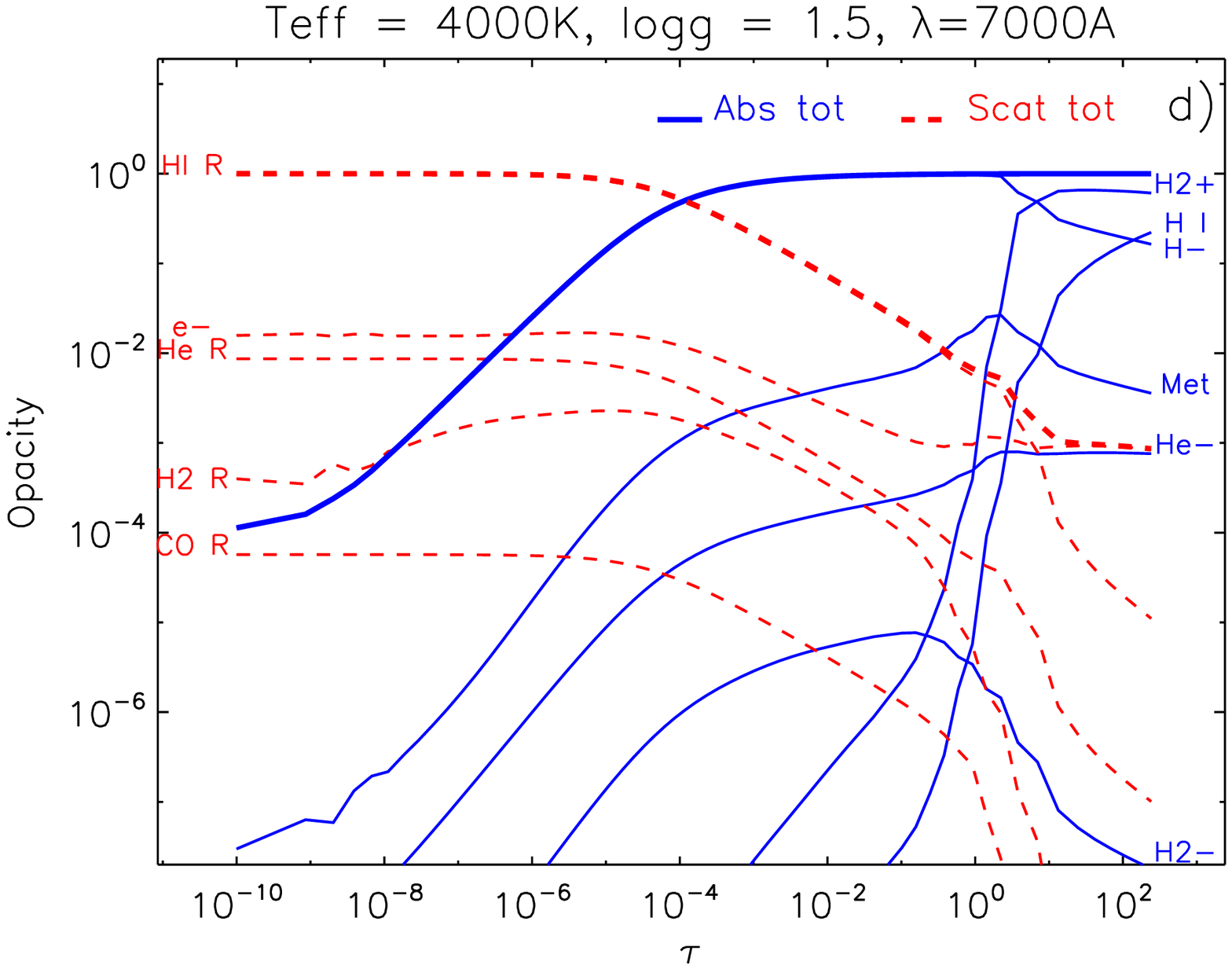}
	\end{minipage} 
 
	\begin{minipage}{0.9\textwidth}
 		 \centering
 		\includegraphics[width=.45\linewidth]{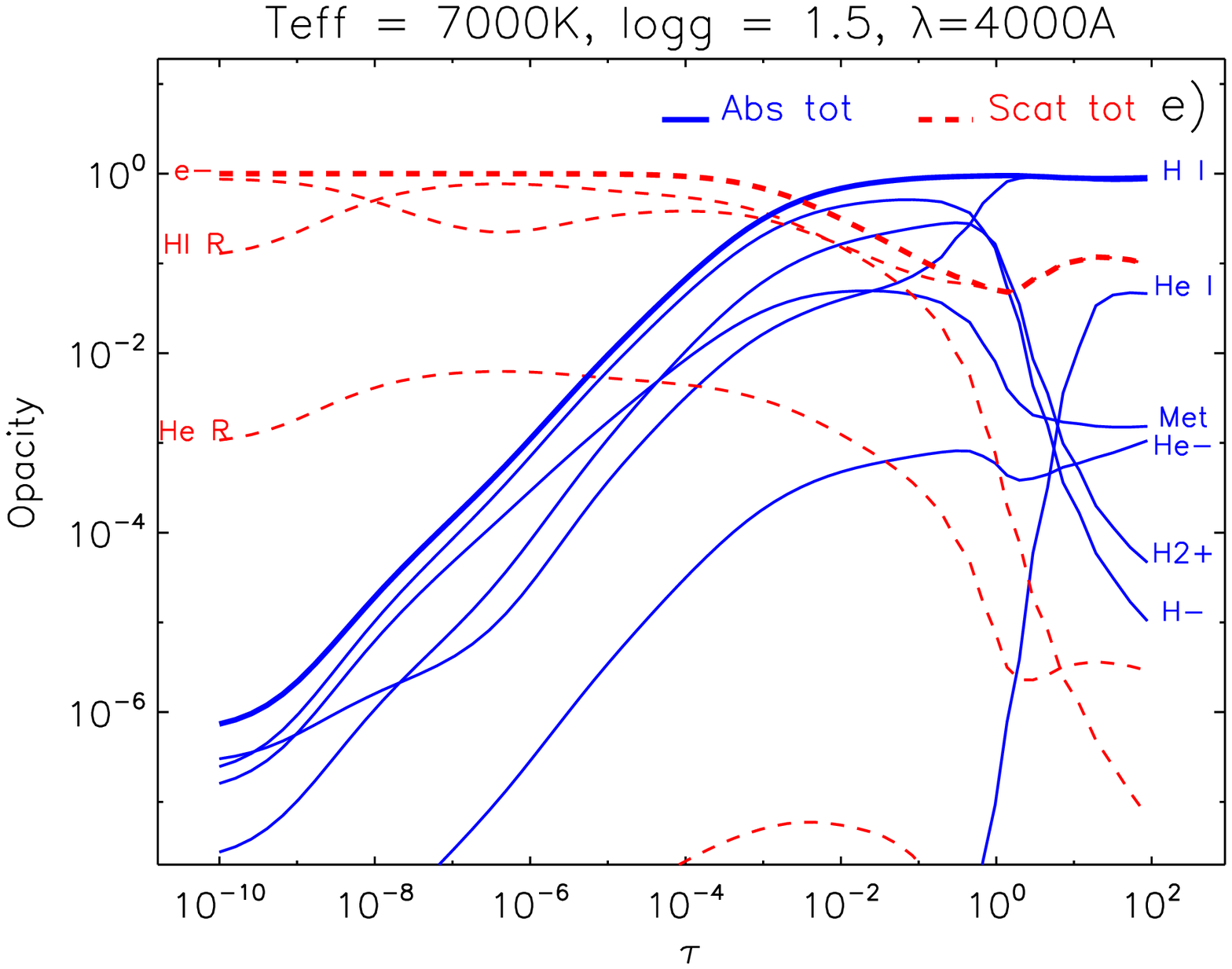}
 		  \includegraphics[width=.45\linewidth]{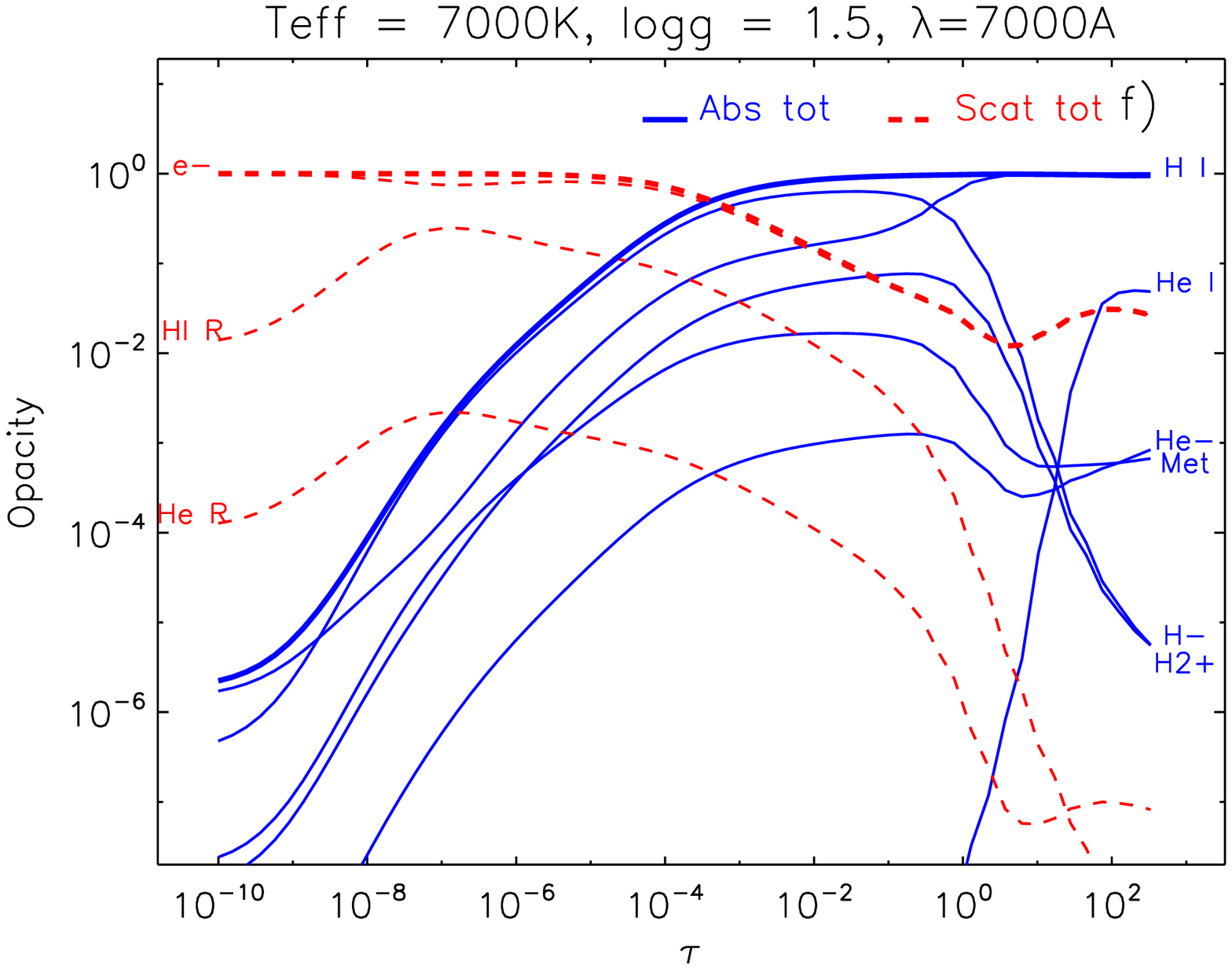}
	\end{minipage}  
\caption{Normalized opacities as a function of optical depth. Titles on each of the panels describe 
the model parameters for the calculation of the opacities. Solid lines show all important absorption 
opacities and dashed lines represent all scattering opacities. Thick solid and thick dashed lines are the 
normalized total absorption and total scattering opacities, respectively.}   
   \label{Fig_opac}
   \end{figure*}

Similarly to \cite{kostogryz15a}, we calculate the scattering and absorption opacities for different wavelengths
using SLOC code \citep{berdyugina91}. The main contributions to continuum opacities for several representative stellar model 
atmospheres are presented in Fig.\ref{Fig_opac}. We consider contributions from: Thomson 
scattering on free electrons $\rm e^-$ and Rayleigh scattering on $\rm HI, HeI, H_2, CO, H_2O$, 
and other molecules (thin dashed lines in Fig.\ref{Fig_opac}).
For calculating absorption opacities we take into account free-free $(\rm ff)$  and bound-free $(\rm bf)$
transitions in $\rm H^-, HI, HeI, He^-, H_2^-, H_2^+$, and metal photoionization(thin solid lines in Fig.\ref{Fig_opac}). 
The thick solid line shows the normalized total absorption and thick dashed 
line the normalized total scattering opacity. The optical depth scale 
is computed using the total opacity at a given wavelength.   

As it is seen from Fig.\ref{Fig_opac}a,b,c,d at upper layer of the atmosphere, absorption by the negative ion hydrogen $\rm H^-$ 
gives the greatest contribution to the opacity at visible wavelengths. In the deeper layers
(for $\tau \ge 1$), absorption by $\rm H^-$ is perhaps not the most dominant but still important contributor 
for solar type and cooler stars. For the hotter giant stars (Fig. \ref{Fig_opac}e,f) 
absorption by $\rm H^-$ is essential in the upper layers of the stellar atmosphere, while 
deeper in the atmosphere ($\tau \ge 1$) it decreases and becomes almost negligible.      
Another absorption opacity source that plays an important role in the deeper layers of 
atmosphere for solar type and cooler stars is $\rm H_2^+$. For hotter giant stars, absorption by $\rm H_2^+$ is 
negligible and the most important opacity source is neutral hydrogen $\rm H I$.   
Note that these absorption opacities do not produce any polarization.

The main contributions to the polarization of the stellar continuum spectrum are scattering sources:
Rayleigh scattering on neutral hydrogen $\rm HI$ 
\citep{chandr60} and on molecular hydrogen $\rm H_2$. The latter is more 
important in cooler atmosphere of dwarf stars (Fig. \ref{Fig_opac}b). For cool giant stars the scattering
by $\rm H_2$ turns out to be less influential (Fig. \ref{Fig_opac}c,d).
Instead, Thomson scattering by free electrons $\rm e^-$ becomes more significant and even dominates in the hot 
and low-gravity stellar atmospheres (Fig. \ref{Fig_opac}e,f).

The behavior of the total scattering (thick dashed line) and absorption (thick solid line) opacity 
is convenient for qualitative understanding the amount of scattering polarization in the atmosphere. 
The deeper is the 'critical' optical depth in the atmosphere where the scattering becomes dominant over the absorption, 
the higher linear polarization is to be expected. Therefore, according 
to Fig.\ref{Fig_opac}, for the stellar atmosphere with $\rm{T_{\rm eff} = 4000 \rm K}$, $\log g = 1.5$ and for the
wavelength $4000 \rm \AA$, where the scattering dominates at $\tau \le 7\times 10^{-1}$ (Fig.\ref{Fig_opac}c), the polarization is 
expected to be very high, as well as for $T_{eff} = 7000 \rm K, \log g = 1.5$ and for all wavelengths. 
On the contrary, for solar-type models with  $\rm{T_{\rm eff} = 5800 \rm K}$, $\log g = 4.5$ at the same wavelength 
the critical height between scattering and absorption is about $\tau \approx 10^{-5}$ that should lead to 
smaller amount of polarized light. We will see that this is indeed the case (see Sect. 3.3).

\section{Results}

We have independently (NK and IM) developed a spherical radiative 
transfer code to compute center-to-limb variation of intensity and scattering 
polarization in the continuum. Both codes 
are based on the iterative solution of the scattering problem but use two 
different methods for the numerical formal solution: Feautrier and short 
characteristics, described in the previous section. The codes usually 
agree down to the relative difference of few hundredth. In the following 
subsections we comment further on the differences. We also note 
that all the results in the paper have been double-checked using both codes.

\subsection{Test of our codes}
\subsubsection{Model atmosphere with equal absorption and scattering opacity }

In this subsection we describe the test of both our approaches to the 
formal solution presented above. We consider an isothermal spherical atmosphere
with inner and outer boundaries at 1 and 30 stellar radii, respectively, as described in \cite{avrettloeser84}.
The total opacity ($k_c + \sigma_c$) is given by $C/r^2$, where the constant $C$ is determined by the 
requirement  that the total radial optical depth of the atmosphere is equal to 4. This results in $C=120/29$, 
and then radial optical depth is given by:
\begin{equation}
	\tau= \frac{120}{29}\Bigg(\frac{1}{r}-\frac{1}{30}\Bigg)
\end{equation}

We assume a constant monochromatic scattering coefficient $\sigma_c = 0.5$. Then, it follows that 
$k_c \equiv 1- \sigma_c = 0.5$. For the given model atmosphere, i.e. variation of temperature, opacity
and scattering coefficient with radius (note that, contrary to the plane-parallel case where $\tau$ is 
chosen as an independent coordinate, we actually require the geometrical scale expressed in physical units), 
we can self-consistently solve the scattering problem and obtain values of the polarized source function at 
all points in the atmosphere as well as the angular variation of emergent polarized intensity 
($I$ and $Q$). There are no available solutions of this specific test case for the polarized radiation so we  start 
by comparing the mean intensity defined as:   
\begin{equation}
	J = \int I(\mu) d\mu, 
\end{equation}
between the two codes and also with other results found in the literature.

\begin{figure}
   	\centering
  	\includegraphics[width=\hsize]{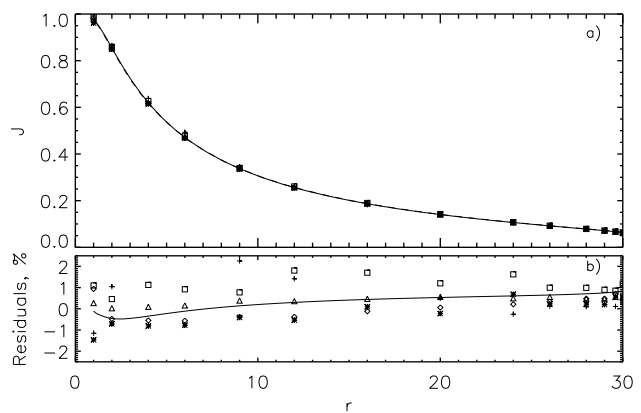}
  \caption{Averaged intensity variation (a) and residuals (relative percentage difference as compared 
      to the calculated with Feautrier method) comparing to those calculated by Feautrier method (b)
      in different layers in the test model atmosphere.
      Solid and dashed lines in (a) correspond to our calculations with Feautrier and short characteristics
      solution of radiative transfer equation. The solid line in (b) describes the difference between Feautrier 
      and short characteristics solutions. Different symbols correspond to values of J(r) obtained
      by different authors: (crosses) are from \cite{mihalas75}, (stars) - from \cite{rogers82}, (squares) - 
     from  \cite{avrettloeser84}, (diamonds) - from \cite{gros97}, (triangles) - from \cite{atanakovich03}.
       }
   \label{Fig_Jr}
\end{figure}
  
The dependence $J(r)$ is well-known for this test case of the spherical atmosphere
\citep{mihalas75, rogers82, avrettloeser84, gros97, atanakovich03}.  
Figure \ref{Fig_Jr}a presents the results of calculations by Feautrier method (solid lines) and 
short characteristics (dashed lines) as compared to other studies:  \cite{mihalas75}(crosses), \cite{rogers82}(stars),  
\cite{avrettloeser84}(squares), \cite{gros97}(diamonds), \cite{atanakovich03}(triangles).
  
As it is difficult to see the difference between calculations, in Fig.\ref{Fig_Jr}b
we present the relative percentage difference (residuals) between those calculated with Feautrier approach ($J^F$) and others (J) in $\%$ obtained as 
\begin{equation}
	\rm {Residuals} = \frac{J - J^{F}}{J^{F}} \times 100 \%. 
\label{residual}
\end{equation}

 The discrepancies for all results are about $1-2 \%$ for different cases (different symbols) 
that can be explained by different discretization in $r$  and different approximations for source function
behavior between any two depth points. Solid line in Fig. \ref{Fig_Jr}b 
describes the residuals between our two codes, which are calculated according 
to  Eq. \ref{residual}, where $J \equiv J^{SC}$ is the solution with short characteristics method, 
which is about $0.5 \%$ of most. Thus, we can conclude that for the test model atmosphere, our calculations agree
within $0.5 \%$.

\begin{figure*}
   \centering
  \includegraphics[width=0.7\hsize]{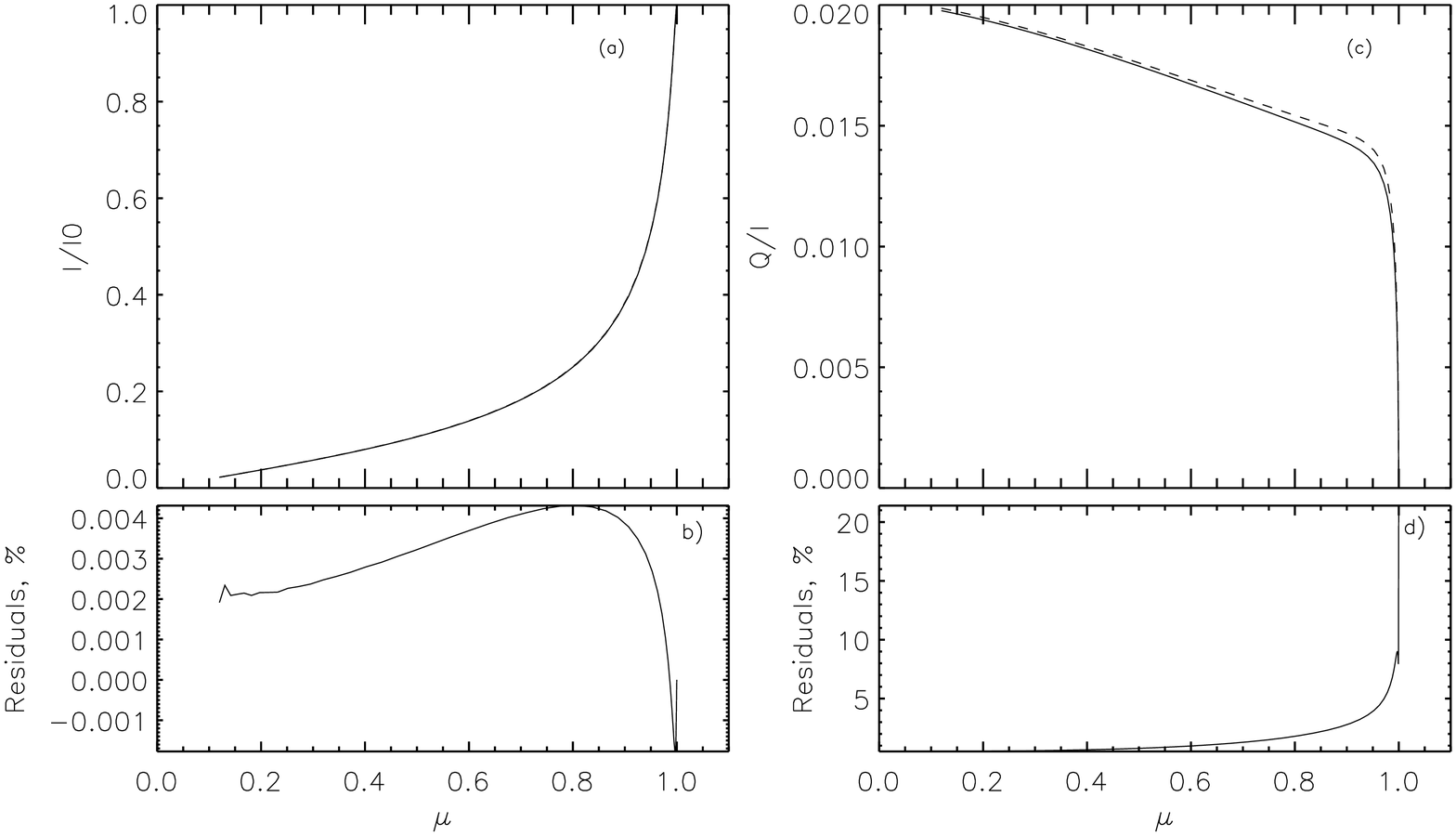}
      \caption{Center-to-limb variation of intensity (a) and polarization (c) for the test model atmosphere. 
      The solid line corresponds to the formal solution of differential radiative transfer equations for polarized light using 
      modified Feautrier method for spherical geometry and the dashed line shows the formal solution of integral
      radiative transfer equation using short characteristics method. The residuals between two approaches
      are depicted in (b) and (d) for center-to-limb variation for intensity and polarization, respectively.
       }
   \label{Fig_comp_test}
   \end{figure*}

In addition to the mean intensity, we are interested in the relative difference 
between our two approaches when computing CLVI and CLVP for this test 
model atmosphere. The comparison is presented in Fig.\,\ref{Fig_comp_test}a,c. 
The results obtained with Feautrier technique are shown with solid lines, and 
the one obtained with short characteristics method are presented by dashed line. The residuals both for 
CLVI and CLVP is given in Fig.\,\ref{Fig_comp_test}b,d. The differences between our codes 
for the intensity are very small (order of percent). The relative difference for the polarization is also very small 
for the major part of the $Q/I (\mu)$ curve. It is only for $\mu$ very close to 1 that our
codes show more significant difference, up to $20 \%$. This is due to the different 
nature of the formal solution: short characteristics method using Bezier splines 
imposes monotonicity of the interpolant (in this case the source function). This 
approach suppresses numerical overshooting but can potentially decrease the accuracy. 
In the Feautrier formal solution, on the other hand, it is implicitly assumed that the 
source function behaves as the second order polynomial. The difference between 
the two approaches is especially prominent in the cases where the source function 
varies a lot between the adjacent depth points which is the case here. Namely, for the ray with 
emergent $\mu$ close to (but not equal to) one, 'local' $\mu$ changes significantly 
along the ray path. This leads to the significant variation of the polarized source function $S_Q$ 
as it strongly depends on the scattering angle. This, in turn, leads to different 
results for the emergent intensity after the formal integration. However, these differences 
become important only for the rays with $\mu \rightarrow 1$, which are of no 
interest for our investigation as the polarization there is very small. In the remainder of the paper, we 
compare the results of the two codes for the realistic model atmospheres.
We found that the relative differences are much lower, down to a few percent. 
 
\subsubsection{Purely scattering atmosphere}

Similar to our previous study \citep{kostogryz15a} as a test of our codes we investigate a center-to-limb
variation of linear polarization in the case of a purely scattering atmosphere. We set the scattering coefficient 
equal to the total opacity, while the absorption coefficient is zero. After such assumptions, the 
intensity and polarization of the outgoing continuum radiation become independent of the frequency and of all 
thermodynamic properties and, therefore, any initial plane-parallel atmosphere can be used.

\begin{figure}
  \centering
  \includegraphics[width=\hsize]{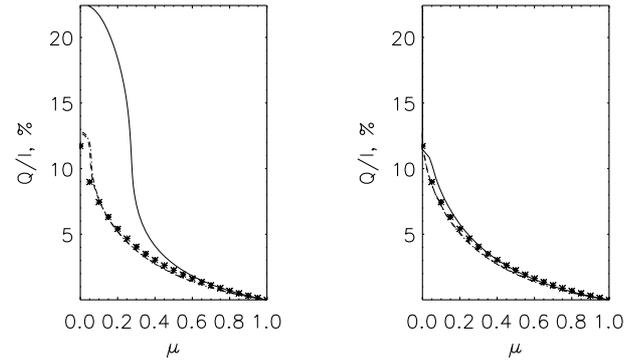}
      \caption{Center-to-limb variation of polarization for the pure scattering atmosphere. Solid line is obtained for
      stellar model with $T_{\rm eff} = 5800 \rm K$ and $\log g = 1.5$ at $4000 \rm \AA$. Two other curves (dashed
      and dotted-dash) are for the models with the same $\log g = 4.5$ but different effective temperatures, 
      $4000 \rm K$ and $5800 \rm K$ at $4000 \rm \AA$ and $7000 \rm \AA$, respectively. The star symbols
      describe the solution for plane-parallel atmosphere obtained by \cite{chandr60}.
      }
   \label{Fig_psc}
   \end{figure}

According to \cite{chandr60}, who
solved the radiative transfer equation for an ideal purely scattering atmosphere in the plane-parallel approximation, 
the polarization increases from the center to the limb and can reach up to $11.7 \%$. Later, \cite{peraiah75} 
solved the radiative transfer equation in spherically symmetric purely scattering homogeneous medium for different ratios 
of extended atmosphere radius to the radius of stellar "surface" and showed 
that for the spherical atmosphere the CLVP can be higher than for plane-parallel atmosphere.
We solve the radiative transfer equation for several Phoenix stellar model atmospheres with effective temperatures 
of $4000 \rm K$ and $5800 \rm K$, $\log g$ of 4.5, and 1.5 and for two different wavelengths of 4000 and $7000 \rm \AA$, 
in which the scattering and absorption coefficients are redefined to get a purely scattering atmosphere.  In the left panel of Fig. \ref{Fig_psc}
 the CLVP for different models with $T_{\rm eff} = 5800 \rm K$, $\log g = 4.5$ at $4000 \rm \AA$ (dashed line), 
 $T_{\rm eff} = 4000 \rm K$, $\log g = 4.5$ at $7000 \rm \AA$ (dash-dotted line) and $T_{\rm eff} = 5800 \rm K$, $\log g = 1.5$ 
 at $4000 \rm \AA$ (solid line) are shown. The star symbols describe the exact solution for a purely scattering plane-parallel atmosphere \citep{chandr60}.
 For stars which do not have extended atmospheres (e.g., solar type stars), we can detect small deviation 
 of CLVP from plane-parallel atmosphere only at the limb, while extended atmospheres show much higher polarization. 
 Here, we notice that the scattering polarization emerging from a purely scattering spherical atmopshere depends on the extension 
 of the atmosphere. For a more extended atmosphere, i.e. smaller $\log g$, we obtain larger differences with respect to plane-parallel geometry.
 In particular the polarization is higher, which agrees with \cite{peraiah75}. 
 
 Following the solar terminology, the limb position of the star is defined as a line of sight where the largest gradient of CLVI occurs. 
 Note that the limb position can only be defined using the solution of radiative transfer equation in a spherical atmosphere.
 Therefore, to explain limb darkening measurements we need to find the limb position and recalculate all $\mu$
  taking into account that $\mu$ at the limb should be equal to zero (i.e. we make a difference between "geometrical" and "observational" $\mu$). 
  In the right panel of Fig. \ref{Fig_psc} 
  we present the same curves but for recalculated values of $\mu$. In this panel we still see small difference
  of a theoretical extended atmosphere (solid line) from Chandrasekhar's solution, while dwarf star atmospheres with different effective temperatures
  and for different wavelengths show very good agreement with the exact solution for plane-parallel atmosphere \citep{chandr60}. This shows
  that scattering has been correctly calculated in the codes.

\subsection{Solar limb darkening and limb polarization}

Direct measurements of center-to-limb variation of intensity (CLVI) and polarization (CLVP)
have been obtained only for the Sun. Therefore, in this subsection we are going to
 investigate how our calculations fit the solar photometric and polarimetric observations. 

\begin{figure*}
   \centering
  \includegraphics[width=0.7\hsize]{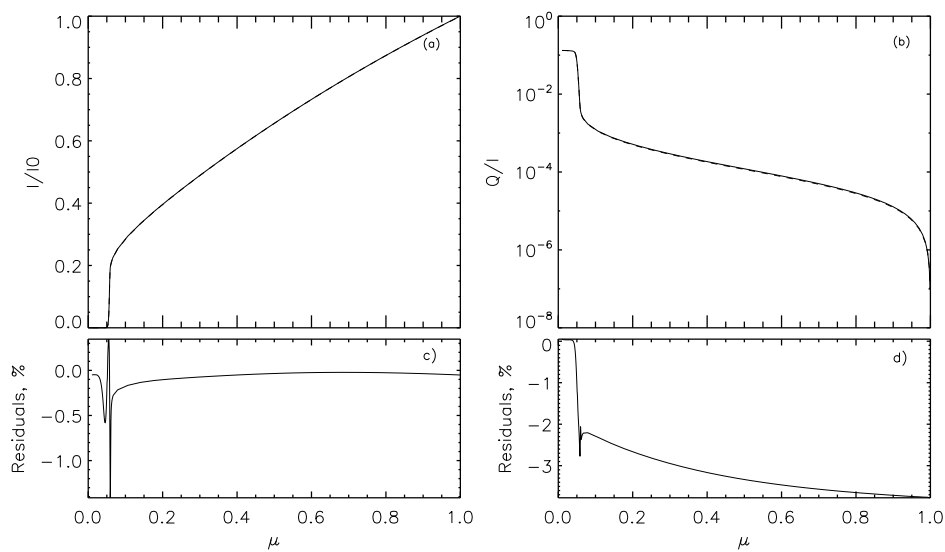}
      \caption{The same as in Fig.\ref{Fig_comp_test} but for Phoenix model atmosphere
      with $T_{\rm eff} = 5800 \rm K, \log g = 4.5$ and wavelength $\lambda = 4500 \rm \AA$. 
      }
  \label{Fig_comp_solar}
\end{figure*}

Before exploring how our calculations fit the observations, the difference 
between our two codes is studied for the solar atmosphere to estimate the difference
in numerical calculations. As is seen from Fig. \ref{Fig_comp_solar} the correlation of our 
calculations are more than sufficient. The relative difference for intensity  (Fig. \ref{Fig_comp_solar}c)
is less then $0.1 - 0.2 \%$ except for the place with residuals equal to $1 \%$, where the intensity gradually decreases 
and defines the limb position. The part of the calculated CLVI for values of $\mu$ outside the limb, are not needed for fitting the measurements.
 For CLVP 
(Fig.\ref{Fig_comp_solar}b,d) we also have very close agreement (residuals are about $3 \%$) for the very small 
values of polarization. Note that from polarimetric solar disk measurements, the error at 
$\mu = 0.1$ is about $15 \%$ and it becomes larger toward the center of the disk and smaller 
toward the limb of the Sun \citep{wb03}. Therefore, we can conclude that the difference between our 
two approaches for solving the radiative transfer equation is negligible for solar-type stars.

\begin{figure*}
   \centering
  \includegraphics[width=\hsize]{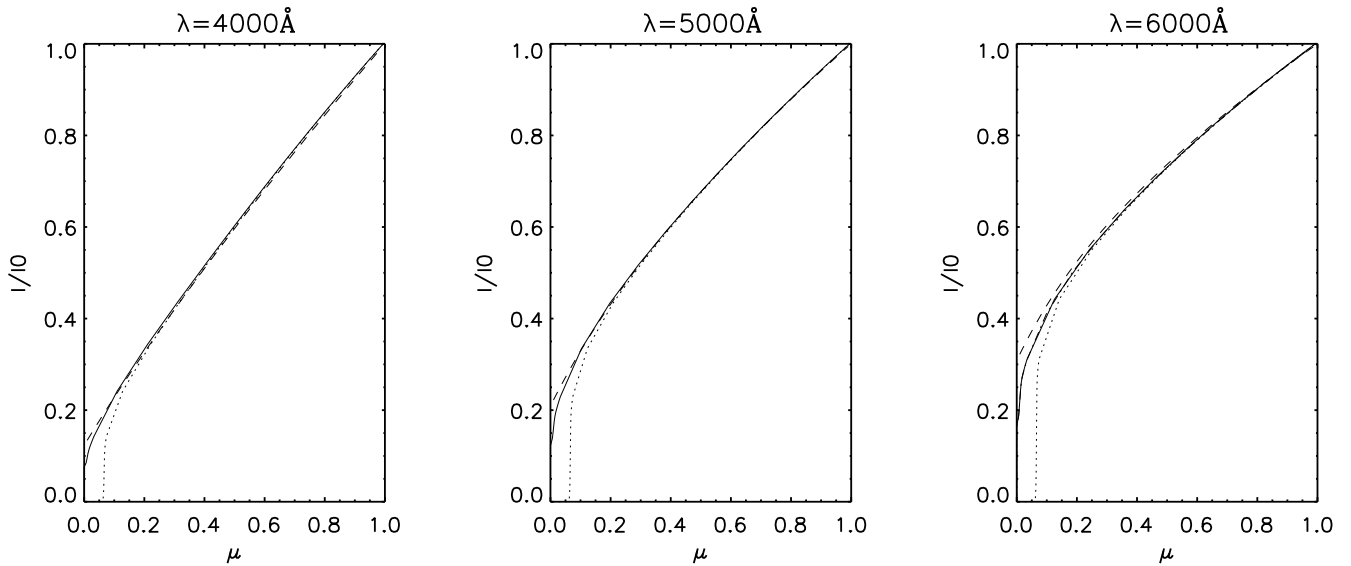}
      \caption{Center-to-limb variation of intensity for the Sun. 
      Solid lines (shifted on $\mu$ according to the limb position) and dotted line (not-shifted) correspond 
      to the formal solution of differential radiative transfer equations for polarized light using the
      modified Feautrier method for spherical geometry with the HSRA solar model atmosphere. Dashed line shows 
      the interpolation of observations by \cite{neckellabs94}
       }
   \label{Fig_obsclvi}
\end{figure*}

A comparison between observations and calculations for CLVI 
and CLVP was already made by \cite{kostogryz15a}, 
where the plane-parallel approximation for model atmosphere was assumed. 
\cite{kostogryz15a} employed the semi-empirical model atmosphere HSRA (averaged quiet Sun) derived
 by \cite{gingerich71} and showed that within some natural brightness variations the simulation for CLVI
fit the observations very well. In this study we use the same observations, which were obtained by \cite{neckellabs94} and 
presented in Fig. \ref{Fig_obsclvi} (dashed line) as analytical polynomial function $P_5(\mu)$. 
This function was obtained by fitting it to the continuum measurements and corrected for scattered light. 
We solve the radiative transfer equation considering spherical symmetry 
for HSRA in order to investigate the matching of our calculation with observations. 
In  Fig. \ref{Fig_obsclvi} we present two calculated curves of CLVI for the different wavelengths 
4000, 5000, and $6000 \rm \AA$ that describe the CLVI with initial (dotted lines) and
recalculated $\mu$ according to the determined limb position on the solar disk (solid lines). 
We show that our calculation (solid line) represents the
observation (dashed line) very well for $5000 \rm \AA$, while there are small discrepancies for $4000 \rm \AA$ and $6000 \rm \AA$
that can be associated either with natural variations of the solar brightness or with imperfect solar 
model atmosphere or with polynomial interpolation of the observations. 
We conclude that the CLVI for both plane-parallel and 
spherical atmospheres can be used for interpretating solar observations. The advantage of taking 
into account a spherical geometry appears at the very limb, where it describes the edge of the solar disk better and thus
provides the information about the solar radius.   

\begin{figure}
   \centering
  \includegraphics[width=\hsize]{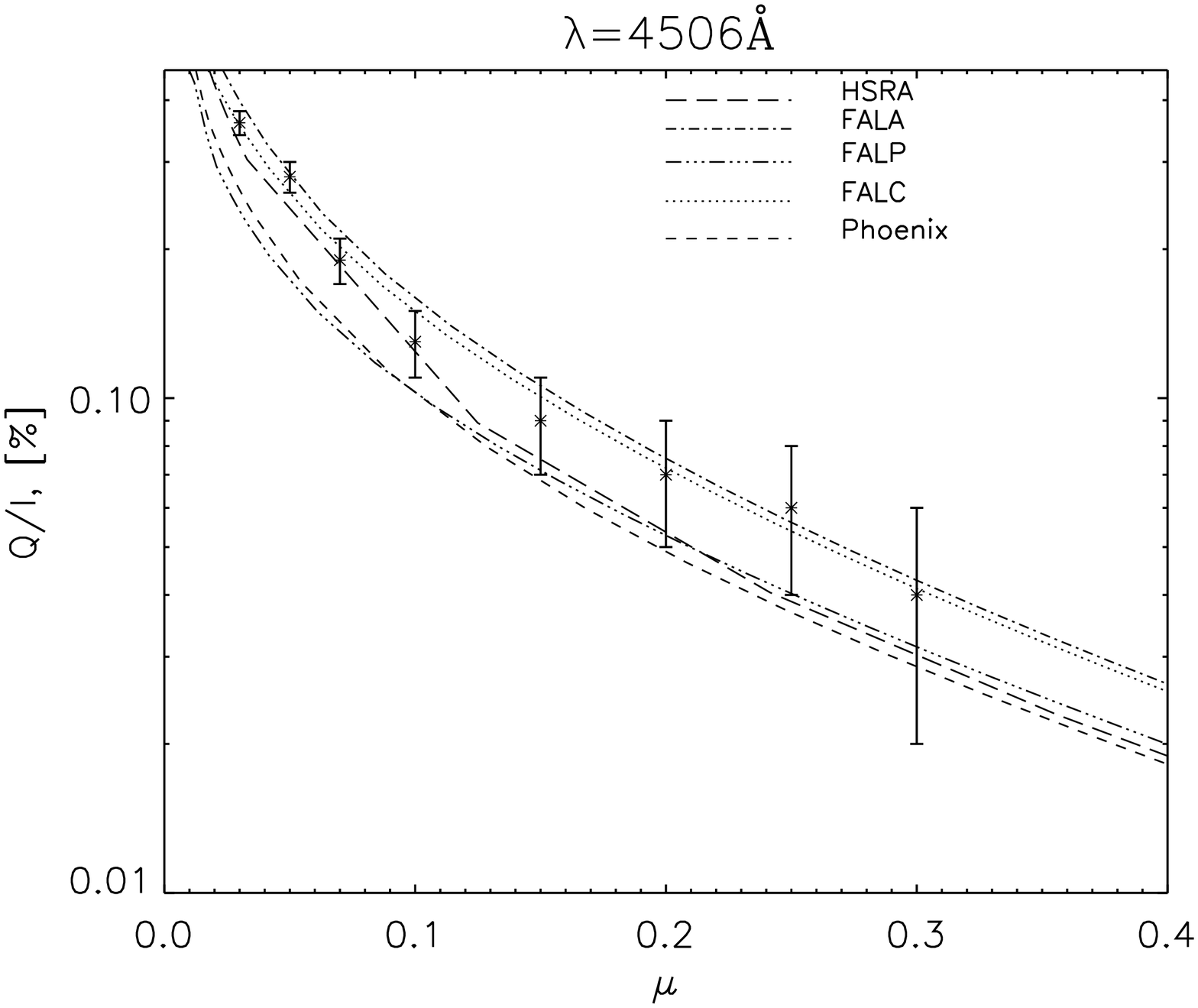}
      \caption{Center-to-limb variation of polarization at $4506 \AA$ for different solar 
      model atmospheres, which are depicted by different line types labeled on the plot.
      The stars with errorbars depict the measurements obtained by \cite{wb03}. The Phoenix 
      model has parameters that are the closest to the solar one, such as $T_{\rm eff} = 5800 \rm K$ 
      and $\log g = 4.5$.
       }
   \label{Fig_obs}
\end{figure}

In addition to CLVI, we calculate center-to-limb variation of continuum polarization for different solar 
model atmospheres of the quiet Sun, such as FALC (averaged quiet Sun), FALA (the supergranular cell center), FALP (the 
plage model) obtained by \cite{fontenla93}, HSRA (averaged quiet Sun) and the spherical PHOENIX model for $T_{\rm eff} = 5800 \rm K$, $\log g = 4.5$ 
 for various wavelengths (4000, 5000, and $6000 \rm \AA$). 
The CLVP calculated for plane-parallel approximation showed very close correspondence with different observations in continuum \citep{kostogryz15a}.
Here we compare the calculated CLVP for spherical geometry with polarimetric 
measurements in continuum obtained by \cite{wb03} (Fig. \ref{Fig_obs}) in order to investigate
the possibility of interpretation of CLVP with spherical atmosphere. As is seen from Fig. \ref{Fig_obs}, 
the theoretical CLVPs for different model atmospheres can describe the behavior of measured 
polarization and the best fit is found for HSRA at a particular wavelength.
We want to mention here that the variation of polarization for different model atmosphere in a spherical geometry
is more significant than in a plane-parallel one. It means that calculated CLVP with spherical geometry
is more sensitive to the solar model atmosphere and can be used for model verification.

\subsection {Stellar limb darkening and polarization}

\begin{figure*}
   \centering
  \includegraphics[width=\hsize]{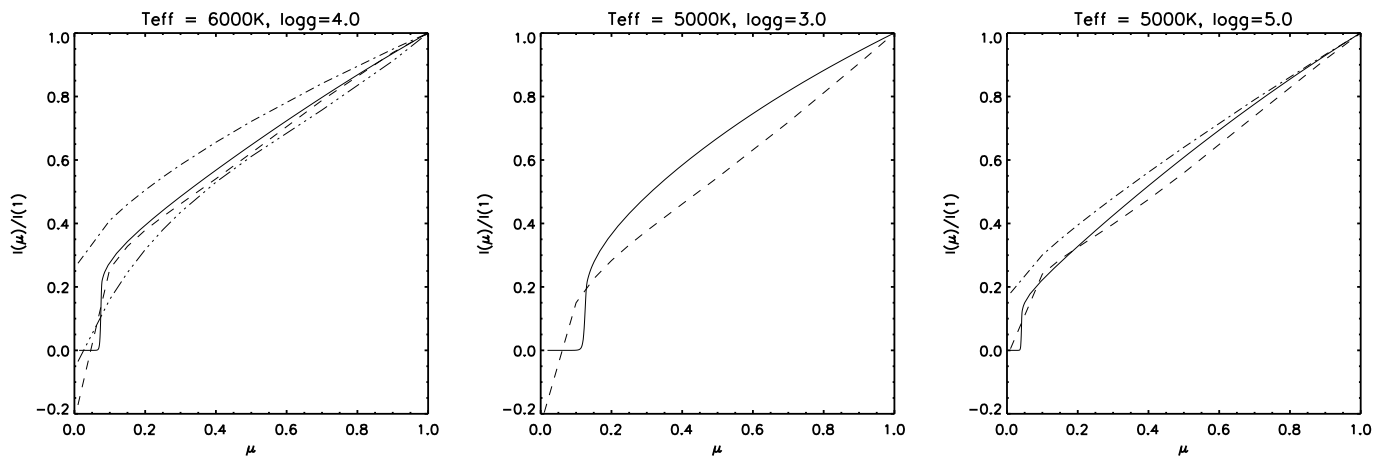}
      \caption{Center-to-limb variation of intensity calculated in the B filter by \cite{claret13} 
      (dashed lines), \cite{neilson13a} (dash-dotted line), \cite{magic15} (dash-triple dotted line) as well as our 
      continuum computation for $4500 \rm \AA$ (solid lines). Different panels correspond to different 
      effective temperatures and surface gravities that are labeled in the title of each plot.
       }
   \label{Fig_LDall}
\end{figure*}

Stellar center-to-limb variation of intensity is studied by different authors in spherical 1D model
atmosphere for different spectral bands and considering different stellar models \citep{claret12, claret13, neilson13b, neilson13a}
and also in 3D plane-parallel model atmosphere \citep{magic15}. We compare our single
wavelength ($4500 \rm \AA$) continuum computation for Phoenix spherical stellar models with other studies for the B filter (Fig. \ref{Fig_LDall}).   
Note that the filter includes contributions from many spectral lines, especially for cooler atmospheres, so we don't
expect a perfect match.

As is seen from the first panel in Fig.\ref{Fig_LDall}, where four various limb darkening curves are presented, 
the different stellar atmospheric models and different authors present diverse results that can be used for testing 
stellar model atmospheres. As expected for hot stars we can reproduce broadband simulations well since 
there are not many spectral lines and the continuum dominates.
At the very limb the curve from \cite{claret13}(solid line) has artificial negative limb darkening that 
may come from the fitting with 4-parameters and not from calculations. 

Because the various stellar models provide different results, in the second and the third panels in Fig. \ref{Fig_LDall} we compare 
our calculations with those from non-linear limb darkening law \citep{claret13} for the same input spherical Phoenix stellar models.
In our previous study of CLVI with plane parallel approximation \citep{kostogryz15a} we discussed that 
the model atmosphere with the same surface gravity and with lower temperature than $5800 \rm K$ that showed a larger deviation from \cite{claret13}. However,
analyzing the CLVI calculated for different surface gravities, we can also add to the previous conclusion that
low gravity star show a larger discrepancy between CLVI in the continuum at $4500 \rm \AA$  and the one calculated in 
the broadband B filter. Naturally, to explain broadband observations of the CLVI we need to take all the spectral lines contributing 
to the passband of the filter into account, while our calculations are useful for explaining the monochromatic measurements
at the continuum level.
 
In order to estimate the maximum difference between our two numerical approaches, we solve the radiative transfer
equation for the most extended model atmosphere with $\log g = 1.0$, effective temperature $4000 \rm K$ at
wavelength $4000 \rm \AA$. 
%
%\begin{figure*}
 %  \centering
 % \includegraphics[width=0.7\hsize]{Figures/Fig_testmodel_stellar.eps}
%      \caption{The same as in Fig.\ref{Fig_comp_test} but for Phoenix model atmosphere
  %    with $T_{\rm eff} = 4000 \rm K, \log g = 1.0$ and wavelength $\lambda = 4000 \rm \AA$. 
%      }
 %\label{Fig_comp_stellar}
%\end{figure*}
%
%As is seen in Fig.\ref{Fig_comp_stellar}, 
The relative difference between two our approaches is about $1.0 - 1.5 \%$ in intensity 
and about $2.0 - 3.0 \%$ in polarization that is smaller than the observational error bars, which can be $10 \%$ or more even 
for solar measurements \citep{wb03}.

\begin{figure}
   \centering
  \includegraphics[width=0.7\hsize]{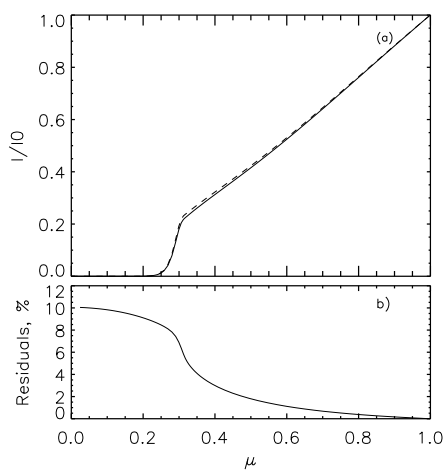}
      \caption {(a) Center-to-limb variation of intensity for Phoenix model atmosphere
      with $T_{\rm eff} = 4000 \rm K, \log g = 1.0$ and wavelength $\lambda = 4000 \rm \AA$ 
      with neglecting polarization (dashed line) and with taking it into account (solid line). 
      (b) Relative difference between two curves from panel (a).}
 \label{Fig_reftest}
\end{figure}

For plane-parallel atmospheres, the center to limb variation of intensity, assuming isotropic scattering, is almost identical to CLVI from a solution with dipole scattering (i.e. one including polarization). In order to check if this holds for extended spherical atmospheres, we calculate CLVI for isotropic and anisotropic scattering, that is, excluding and including polarization, respectively.
In Figure \ref{Fig_reftest} we present the results of our calculations 
for Phoenix model atmosphere with $T_{\rm eff} = 4000 \rm K, \log g = 1.0$ and wavelength $\lambda = 4000 \rm \AA$. 
For this model the relative difference between these two calculation for CLVI is about $8\%$ at the limb, while for more compact 
stellar model atmospheres this effect is smaller.
   
We solve the radiative transfer equation for polarized light taking spherical 
symmetry of a star into account for the grid of spherical Phoenix model atmospheres within the range of effective 
temperatures from 4000 to $7000 \rm K$ at steps of $100 \rm K$  and for $\log g$ from 1.0 to 5.5 at steps of 0.5.
For  different wavelengths, effective temperatures, $\log g$ and various position
on the stellar disk $\mu$, we present the values of CLVI in Table 1 and of CLVP in Table 2. In addition
we present $\mu_{limb}$ that is equal to the calculated $\mu$ where the largest gradient in the limb darkening
curve occurs and corresponds to the real radius of a star, which is calculated and presented in Table 1 as well.
This is important when interpreting exoplanet transit data which provide only a planet-to-star radii ratio.

\begin{sidewaystable*}
\caption{\label{CLVI} Calculated center-to-limb variation of intensity for different stellar parameters. All values of $I(\mu)/I(1.0) = 1.0$ at $\mu=1.0$. The complete table is available at the CDS.}
\centering
\begin{tabular}{c c c c c c c c c c c c c c c}
\hline
\multicolumn{1}{c}{Wavelength, $\AA$} & \multicolumn{1}{c}{T, K} & \multicolumn{1}{c}{$\log g$} &  \multicolumn{1}{c}{$\mu_{limb}$} &\multicolumn{1}{c}{$R_\star/R_\sun$} & \multicolumn{10}{c}{$I(\mu)/I(1.0)$} \\
\multicolumn{1}{c}{}                   & \multicolumn{1}{c}{}       & \multicolumn{1}{c}{}   & \multicolumn{1}{c}{}   & \multicolumn{1}{c}{} 	& 0.01     & 0.04     & 0.09 	& 0.16 & 0.24 & 0.35 & 0.46 & 0.58 & 0.72 & 0.86   \\
\hline
4000 & 4000 & 1.0 & 0.32 & 59.20 & 2.24e-08 & 8.92e-08 & 3.90e-07 & 6.68e-06 & 1.08e-03 & 2.11e-01 & 3.74e-01 & 5.42e-01 & 7.04e-01 & 8.57e-01 \\ 
4000 & 4000 & 1.5 & 0.24 & 34.18 & 4.41e-08 & 1.84e-07 & 1.76e-06 & 1.70e-04 & 1.57e-01 & 2.86e-01 & 4.28e-01 & 5.76e-01 & 7.24e-01 & 8.66e-01 \\ 
4000 & 4000 & 2.0 & 0.20 & 15.87 & 6.37e-08 & 3.19e-07 & 5.78e-06 & 2.56e-03 & 1.86e-01 & 3.04e-01 & 4.37e-01 & 5.78e-01 & 7.22e-01 & 8.63e-01 \\ 
4000 & 4000 & 2.5 & 0.16 & 7.52 & 9.37e-08 & 5.80e-07 & 2.93e-05 & 7.53e-02 & 1.97e-01 & 3.02e-01 & 4.26e-01 & 5.64e-01 & 7.10e-01 & 8.57e-01 \\ 
4000 & 4000 & 3.0 & 0.12 & 4.26 & 1.64e-07 & 1.73e-06 & 8.58e-04 & 1.33e-01 & 2.10e-01 & 3.08e-01 & 4.26e-01 & 5.62e-01 & 7.07e-01 & 8.55e-01 \\ 
4000 & 4000 & 3.5 & 0.09 & 2.22 & 2.56e-07 & 6.25e-06 & 5.16e-02 & 1.41e-01 & 2.13e-01 & 3.06e-01 & 4.21e-01 & 5.54e-01 & 7.00e-01 & 8.50e-01 \\ 
4000 & 4000 & 4.0 & 0.07 & 1.25 & 4.38e-07 & 5.46e-05 & 1.06e-01 & 1.72e-01 & 2.47e-01 & 3.40e-01 & 4.50e-01 & 5.75e-01 & 7.12e-01 & 8.55e-01 \\ 
4000 & 4000 & 4.5 & 0.05 & 0.64 & 7.55e-07 & 7.71e-04 & 1.55e-01 & 2.34e-01 & 3.17e-01 & 4.09e-01 & 5.13e-01 & 6.25e-01 & 7.46e-01 & 8.72e-01 \\ 
4000 & 4000 & 5.0 & 0.04 & 0.36 & 1.58e-06 & 1.03e-01 & 2.24e-01 & 3.14e-01 & 4.05e-01 & 4.99e-01 & 5.96e-01 & 6.96e-01 & 7.98e-01 & 9.00e-01 \\ 
4000 & 4000 & 5.5 & 0.03 & 0.20 & 3.48e-06 & 1.67e-01 & 2.82e-01 & 3.79e-01 & 4.74e-01 & 5.67e-01 & 6.59e-01 & 7.48e-01 & 8.36e-01 & 9.20e-01 \\ 
4000 & 4100 & 1.0 & 0.31 & 60.81 & 2.43e-08 & 9.67e-08 & 4.38e-07 & 7.87e-06 & 1.34e-03 & 2.28e-01 & 4.06e-01 & 5.74e-01 & 7.30e-01 & 8.72e-01 \\ 
4000 & 4100 & 1.5 & 0.23 & 35.06 & 4.59e-08 & 1.93e-07 & 1.92e-06 & 2.02e-04 & 1.67e-01 & 3.06e-01 & 4.52e-01 & 5.99e-01 & 7.40e-01 & 8.75e-01 \\ 
4000 & 4100 & 2.0 & 0.20 & 16.26 & 6.71e-08 & 3.37e-07 & 6.34e-06 & 3.08e-03 & 1.98e-01 & 3.23e-01 & 4.59e-01 & 5.99e-01 & 7.38e-01 & 8.73e-01 \\ 
4000 & 4100 & 2.5 & 0.16 & 7.71 & 9.81e-08 & 6.17e-07 & 3.32e-05 & 8.66e-02 & 2.12e-01 & 3.23e-01 & 4.49e-01 & 5.87e-01 & 7.29e-01 & 8.68e-01 \\ 
4000 & 4100 & 3.0 & 0.12 & 4.36 & 1.71e-07 & 1.84e-06 & 1.01e-03 & 1.42e-01 & 2.25e-01 & 3.27e-01 & 4.49e-01 & 5.83e-01 & 7.24e-01 & 8.64e-01 \\ 
4000 & 4100 & 3.5 & 0.09 & 2.28 & 2.66e-07 & 6.73e-06 & 5.83e-02 & 1.49e-01 & 2.26e-01 & 3.23e-01 & 4.39e-01 & 5.71e-01 & 7.12e-01 & 8.57e-01 \\ 
4000 & 4100 & 4.0 & 0.07 & 1.28 & 4.43e-07 & 5.86e-05 & 1.01e-01 & 1.62e-01 & 2.38e-01 & 3.33e-01 & 4.47e-01 & 5.77e-01 & 7.15e-01 & 8.59e-01 \\ 
4000 & 4100 & 4.5 & 0.05 & 0.66 & 7.47e-07 & 8.32e-04 & 1.34e-01 & 2.05e-01 & 2.86e-01 & 3.79e-01 & 4.86e-01 & 6.05e-01 & 7.33e-01 & 8.66e-01 \\ 
4000 & 4100 & 5.0 & 0.04 & 0.37 & 1.54e-06 & 9.41e-02 & 1.96e-01 & 2.79e-01 & 3.66e-01 & 4.59e-01 & 5.59e-01 & 6.65e-01 & 7.75e-01 & 8.88e-01 \\ 
4000 & 4100 & 5.5 & 0.03 & 0.21 & 3.53e-06 & 1.55e-01 & 2.61e-01 & 3.53e-01 & 4.46e-01 & 5.40e-01 & 6.34e-01 & 7.28e-01 & 8.21e-01 & 9.12e-01 \\ 

\end{tabular}

\vspace{20pt}

\caption{\label{CLVQ} Calculated center-to-limb variation of linear polarization in continuum spectra of different stars. 
All values of $Q/I(\mu) = 0.0$ at $\mu=1.0$. The complete table is available at the CDS.}
\centering
\begin{tabular}{c c c c c c c c c c c c c c}
\hline
\multicolumn{1}{c}{Wavelength, $\AA$} & \multicolumn{1}{c}{T, K} & \multicolumn{1}{c}{$\log g$} & \multicolumn{1}{c}{$\mu_{limb}$} & \multicolumn{10}{c}{$Q/I(\mu)$} \\
\multicolumn{1}{c}{}                   & \multicolumn{1}{c}{}       & \multicolumn{1}{c}{} & \multicolumn{1}{c}{}   & 0.01     & 0.04     & 0.09 & 0.16 & 0.24 & 0.35 & 0.46 & 0.58 & 0.72 & 0.86  \\
\hline

4000 & 4000 & 1.0 & 0.32 & 0.304713 & 0.304713 & 0.300721 & 0.286831 & 0.254027 & 0.109844 & 0.036350 & 0.016516 & 0.007732 & 0.002942\\ 
4000 & 4000 & 1.5 & 0.24 & 0.262117 & 0.261923 & 0.256183 & 0.238953 & 0.145180 & 0.039959 & 0.018404 & 0.009288 & 0.004552 & 0.001771\\ 
4000 & 4000 & 2.0 & 0.20 & 0.245622 & 0.244932 & 0.238526 & 0.217849 & 0.057223 & 0.023136 & 0.011393 & 0.005921 & 0.002945 & 0.001156\\ 
4000 & 4000 & 2.5 & 0.16 & 0.238193 & 0.237209 & 0.229789 & 0.161957 & 0.031961 & 0.014795 & 0.007608 & 0.004003 & 0.001996 & 0.000784\\ 
4000 & 4000 & 3.0 & 0.12 & 0.226966 & 0.225539 & 0.215860 & 0.041905 & 0.017850 & 0.009014 & 0.004790 & 0.002560 & 0.001284 & 0.000505\\ 
4000 & 4000 & 3.5 & 0.09 & 0.222013 & 0.220085 & 0.133558 & 0.022820 & 0.010889 & 0.005689 & 0.003075 & 0.001654 & 0.000831 & 0.000327\\ 
4000 & 4000 & 4.0 & 0.07 & 0.202890 & 0.200179 & 0.026698 & 0.010555 & 0.005377 & 0.002950 & 0.001659 & 0.000918 & 0.000471 & 0.000188\\ 
4000 & 4000 & 4.5 & 0.05 & 0.172316 & 0.166012 & 0.009512 & 0.004160 & 0.002239 & 0.001294 & 0.000762 & 0.000439 & 0.000233 & 0.000095\\ 
4000 & 4000 & 5.0 & 0.04 & 0.140255 & 0.017396 & 0.003375 & 0.001603 & 0.000887 & 0.000525 & 0.000316 & 0.000186 & 0.000100 & 0.000042\\ 
4000 & 4000 & 5.5 & 0.03 & 0.119067 & 0.004580 & 0.001420 & 0.000710 & 0.000394 & 0.000233 & 0.000140 & 0.000082 & 0.000044 & 0.000018\\ 
4000 & 4100 & 1.0 & 0.31 & 0.293314 & 0.293314 & 0.289127 & 0.274888 & 0.241629 & 0.090583 & 0.029629 & 0.013601 & 0.006428 & 0.002471\\ 
4000 & 4100 & 1.5 & 0.23 & 0.253358 & 0.253132 & 0.247301 & 0.229617 & 0.123994 & 0.033501 & 0.015497 & 0.007890 & 0.003906 & 0.001534\\ 
4000 & 4100 & 2.0 & 0.20 & 0.237968 & 0.237263 & 0.230766 & 0.209607 & 0.049004 & 0.019686 & 0.009749 & 0.005100 & 0.002550 & 0.001005\\ 
4000 & 4100 & 2.5 & 0.16 & 0.229820 & 0.228803 & 0.221244 & 0.142061 & 0.027218 & 0.012595 & 0.006518 & 0.003454 & 0.001733 & 0.000685\\ 
4000 & 4100 & 3.0 & 0.12 & 0.218847 & 0.217395 & 0.207480 & 0.036181 & 0.015314 & 0.007740 & 0.004132 & 0.002222 & 0.001123 & 0.000445\\ 
4000 & 4100 & 3.5 & 0.09 & 0.215032 & 0.213075 & 0.118548 & 0.020170 & 0.009538 & 0.004977 & 0.002704 & 0.001465 & 0.000742 & 0.000294\\ 
4000 & 4100 & 4.0 & 0.07 & 0.206371 & 0.203705 & 0.027709 & 0.011076 & 0.005527 & 0.002965 & 0.001637 & 0.000896 & 0.000456 & 0.000181\\ 
4000 & 4100 & 4.5 & 0.05 & 0.184223 & 0.177853 & 0.011312 & 0.004879 & 0.002568 & 0.001450 & 0.000836 & 0.000473 & 0.000247 & 0.000100\\ 
4000 & 4100 & 5.0 & 0.04 & 0.152530 & 0.019017 & 0.003971 & 0.001884 & 0.001038 & 0.000611 & 0.000365 & 0.000214 & 0.000115 & 0.000047\\ 
4000 & 4100 & 5.5 & 0.03 & 0.126389 & 0.004956 & 0.001555 & 0.000779 & 0.000434 & 0.000258 & 0.000155 & 0.000091 & 0.000049 & 0.000021\\ 
\end{tabular}
\end{sidewaystable*}

\begin{figure*}
   \centering
  \includegraphics[width=\hsize]{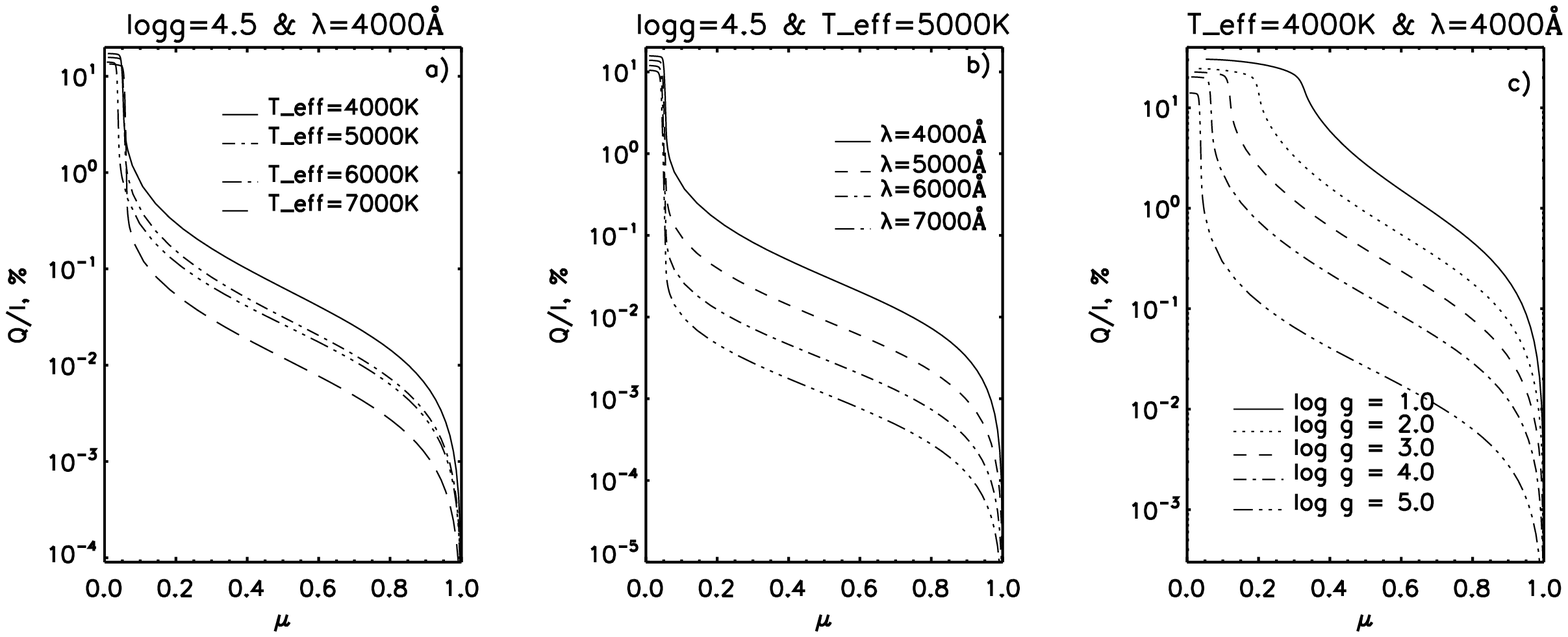}
      \caption{Center-to-limb variation of continuum polarization for different spherical 
      stellar model atmospheres. The title above each of the panels indicates the fixed parameters,
      while labels on each plot describe different curves. Note continuum polarizations
      are given in a logarithmic scale.
       }
    \label{Fig_clvp}
\end{figure*}

In Fig.\ref{Fig_clvp} the dependence of stellar continuum polarization on the spherical model atmosphere with different effective temperatures, 
surface gravities, and wavelengths are presented. For the fixed surface gravity ($\log g = 4.5$) and the wavelength ($\lambda = 4000 \rm \AA$),
give as titles in Fig. \ref{Fig_clvp}a, the cooler atmosphere shows larger CLVP. Fixing the surface gravity and temperature, one 
can investigate the behavior of continuum polarization in different wavelengths. Figure \ref{Fig_clvp}b  shows that at shorter wavelengths 
higher polarization can be obtained. However, there are some cases in our grid of model atmospheres where we have larger polarization for 
longer wavelengths and hotter temperatures. It is true for giant and supergiant stars with $\log g \le 1.5$ where the main 
source of scattering opacity is Thomson scattering on free electrons (see Fig. \ref{Fig_opac}e, f), which is independent on wavelength 
but increasing with increasing temperature. The last panel (Fig. \ref{Fig_clvp}c) shows that the higher gravity of a star leads to lower 
linear polarization. 

\begin{figure}
   \centering
  \includegraphics[width=\hsize]{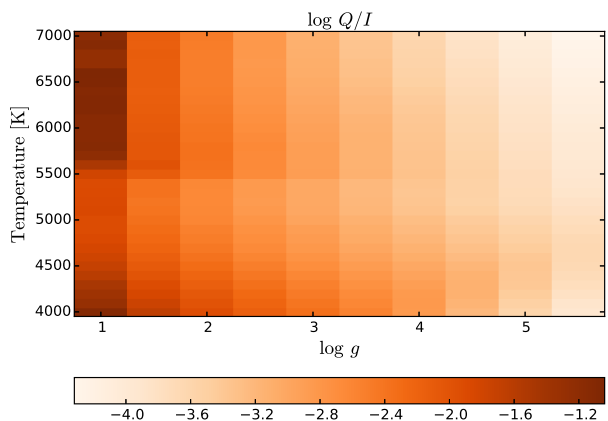}
      \caption{
      Center-to-limb variation of continuum polarization for different 
      stellar models. The color scale shows the logarithm of
      integrated continuum polarization within $\mu$ from 0 to 0.3 at
      wavelength $4000 \rm \AA $.
       }
   \label{Fig_clvp_map}
\end{figure}

In Fig. \ref{Fig_clvp_map} we present the distribution of integrated value of polarization within the range of $\mu = 0.0 - 0.3$
depending on effective temperature and gravity of a star at $\lambda = 4000 \rm \AA$. We 
see that polarization degree depends on the atmosphere extension (i.e. scale height), which varies with the
effective temperatures and surface gravities.
As is seen from Fig. \ref{Fig_clvp_map}, 
the coolest stars with $\log g  = 3.0 - 4.5 $ show larger polarization, while for very compact dwarf stars with $\log g  = 5.0 - 5.5$
the highest polarization is seen for $T_{\rm eff} \approx 4200 - 4600 K$ (depending on $\log g$) and decreases for smaller temperatures. 
For giant and supergiant stars ($\log g < 3.0$), the situation
is different. The limb polarization is the largest for the hottest stars where the Thomson scattering plays an important role in opacity, 
then  it decreases  for the stars with effective temperatures 
of about $4500 - 5500 \rm K$ (depending on $\log g$), and cooler stars again show an increase of 
polarization because of Rayleigh scattering that dominates
in scattering opacity (see, Fig. \ref{Fig_opac}).  For coolest 
and most compact dwarfs limb polarization decreases. This complex behavior can be explained by the interplay of opacity and 
geometry that depends on the variation of physical parameters in the atmosphere, such as effective temperature and density.  

Therefore, we conclude that for the range of effective temperatures of $4000 - 7000 \rm K$ and surface gravities
of 3.0 - 5.0, that were also considered in our previous study \citep{kostogryz15a} with the plane-parallel approximation, 
the continuum polarization is the largest for low-gravity cool stars. The extension of the grid of stellar model
atmospheres, especially to smaller $\log g$, shows the increasing of CLVP for hot stars, where the Thomson scattering
becomes dominating in scattering opacities and also increasing of CLVP for cool stars, where Rayleigh scattering on $\rm HI$ 
is still the most important scattering opacity. For the range of temperatures of $45000 - 5500 \rm K$ (depending on $\log g$) 
the decreasing of limb polarization in continuum occurs.  

\section{Conclusions}
In this paper we developed two independent codes for solving radiative transfer equation of polarized light in continuum considering spherical
stellar atmospheres and presented the center-to-limb variations of intensity and polarization for the range of effective temperatures 
of $4000 - 7000 \rm K$, surface gravities of $\log g = 1.0 - 5.5$, and for several wavelengths 
($4000, 4500, 5000, 6000, 7000 \rm \AA$). We showed that two different formal solutions, 
such as Feautrier method and short characteristics, provide very similar results within small deviations that are discussed in the paper. 

For the test isothermal model atmosphere with inner and outer boundaries at 1 and 30 stellar radii and constant opacity $(\sigma = 0.5)$ 
we calculated the mean intensity and compared our results with the already published values. It was shown that the discrepancies for all results
are about $1 - 2 \%$ that can be explained by different discretizations in $r$ and different approximations of source function. As we have the same 
discretization in $r$ but still different approximations for description of source function, the differences between 
our results is within $0.5 \%$. For this test model the CLVI and CLVP calculations with two different formal solutions were compared and 
showed that the intensity varies within $0.4 \%$, while the deviation in polarization is larger. Anyway, for the most extended real atmosphere
the deviations between the two codes is approximately $2 - 3 \%$.

For the solar atmosphere we obtained very good agreement for CLVI and CLVP with observations. 
Moreover, we compare our previous calculations for a plane-parallel
atmosphere with calculations in a spherical atmosphere and showed that CLVI in a spherical atmosphere is more informative at the limb and 
provides information about the radius of the Sun
The CLVP in a spherical atmosphere for different models have larger deviations than in plane-parallel atmosphere and can be used 
for testing solar model atmospheres.
       
For extended stellar atmospheres (giant and super-giant stars) with $\log g = 1.0 - 2.5$ all models provide large polarization with 
two maxima for cooler and for hotter atmospheres. These two maxima in polarization consist of two different scattering
processes for hot and cool atmospheres. So, Thomson scattering on free electrons are important and even dominating scattering 
process for hot giant stars, while for cool giant stars Rayleigh scattering on $\rm HI$ is the main contributor to total scattering 
opacity source. Naturally, the polarization for cool giant stars is larger for shorter wavelengths, while for hot giant stars it does not depend 
on wavelengths so much. 

In sub-giant and dwarf stellar atmospheres with $\log g = 3.0 - 4.5$, Rayleigh scattering on HI is the most important opacity. Therefore,
the largest polarization can be detected in low-gravity cool star at shorter wavelengths. 

For very compact cool dwarf stars ($\log g = 5.0 - 5.5$) the maximum polarization can be obtained for model
atmospheres within the range of $T_{\rm eff} \approx 4200 - 4600 K$ and it decreases for the cooler and for 
the hotter atmospheres.

The radius of the star for each model atmospheres is tabulated and can be used for interpreting exoplanet 
transit curves as it yields information only about planet-to-star radii ratio.

\begin{acknowledgements}
     This   work   was   supported   by   the   European   Research
Council  Advanced  Grant  HotMol(ERC-2011-AdG291659). IM acknowledges partial support of Serbian Ministry of Education and Science, through the project 176004, "Stellar Physics." We thank 
Marianne Faurobert and Taras Yakobchuk for useful discussions.       
\end{acknowledgements}

%-------------------------------------------------------------------
\bibliographystyle{aa} % style aa.bst
\bibliography{Cont.bib} % your references Yourfile.bib
  
\end{document}